\documentclass[letterpaper,twocolumn,10pt]{article}
\usepackage{mainstyle}

\usepackage{url}
\usepackage{color}
\usepackage{caption}
\usepackage{diagbox}
\usepackage{times}
\usepackage{tikz}
\usepackage{amsmath}
\usepackage{booktabs}
\usepackage{makecell}
\usepackage{multicol}
\usepackage{multirow}
\usepackage{sandlab}
\usepackage{tablefootnote}
\usepackage{subfig}
\usepackage{arydshln}
\usepackage{enumitem}
\usepackage{listings}

\makeatletter
\patchcmd{\maketitle}
  {\@maketitle}
  {\vspace{-0.4in}\@maketitle\vspace{-0.2in}}
  {}
  {}
\makeatother

\newcommand{\para}[1]{{\vspace{3pt} \noindent \textbf{#1}
    \hspace{6pt}}}

\newcommand{\secspace}{\vspace{-0.05in}}

\newcommand\blfootnote[1]{
  \begingroup
  \renewcommand\thefootnote{}\footnote{#1}
  \addtocounter{footnote}{-1}
  \endgroup
}

\newenvironment{packed_itemize}{
\begin{list}{\labelitemi}{\leftmargin=0.5em}
  \setlength{\itemsep}{3pt}
  \setlength{\parskip}{0pt}
  \setlength{\parsep}{0pt}
  \setlength{\headsep}{0pt}
  \setlength{\topskip}{0pt}
  \setlength{\topmargin}{0pt}
  \setlength{\topsep}{0pt}
  \setlength{\partopsep}{0pt}
}{\end{list}}

\lstdefinestyle{mystyle}{
    basicstyle=\ttfamily\footnotesize
}
\begin{document}

\date{}

\title{\Large \bf  Inception Attacks: Immersive Hijacking in Virtual Reality}

\author{Zhuolin Yang, Cathy Yuanchen Li, Arman Bhalla, Ben Y. Zhao,
  Haitao Zheng\\ Department of Computer Science, University of Chicago
} 

\maketitle

\begin{abstract}
Today's virtual reality~(VR) systems provide immersive
interactions that seamlessly connect users with online services and
one another. However, these immersive interfaces also introduce new
vulnerabilities, making it easier for users to fall prey to new attacks. 
In this work, we introduce the {\em immersive hijacking} attack, where
a remote attacker takes control of a user's interaction
with their VR system, by trapping them inside a malicious app that
masquerades as the full VR interface. Once trapped, all of the user's
interactions with apps, services and other users can be recorded and
modified without their knowledge.
 This not only allows traditional privacy attacks but also introduces
 new interaction attacks, where two VR users encounter 
 vastly different immersive experiences during their interaction.  We present our
 implementation of the immersive hijacking attack on Meta Quest headsets and
 conduct IRB-approved user studies that validate its efficacy and
 stealthiness. Finally, we examine effectiveness and tradeoffs of various 
  potential defenses, and propose a multifaceted 
  defense pipeline. \blfootnote{© 2024 Copyright held by the owner/author(s).}
  \blfootnote{This is the author's version of the work. It is posted here for your personal use. Not for redistribution.}
\end{abstract}

\vspace{-0.05in}
\section{Introduction}
\label{sec:intro} \vspace{-0.08in}
Recent advances in virtual reality (VR) are poised to
change the way we interact with the world and each
other~\cite{vr_design, education1,RADIANTI2020103778, vr_finance,
  Sugimoto2022,vr_anesthesia,healthcare1,healthcare2}. VR headsets
offer an increasingly immersive experience
comparable to reality itself,  eliminating geographical barriers and
facilitating social and workplace
interactions through the use of personalized
avatars~\cite{vrchat,vrdating,vr_work, vr_business}.

\begin{figure}[t]
  \centering 
  \includegraphics[width=0.7\linewidth]{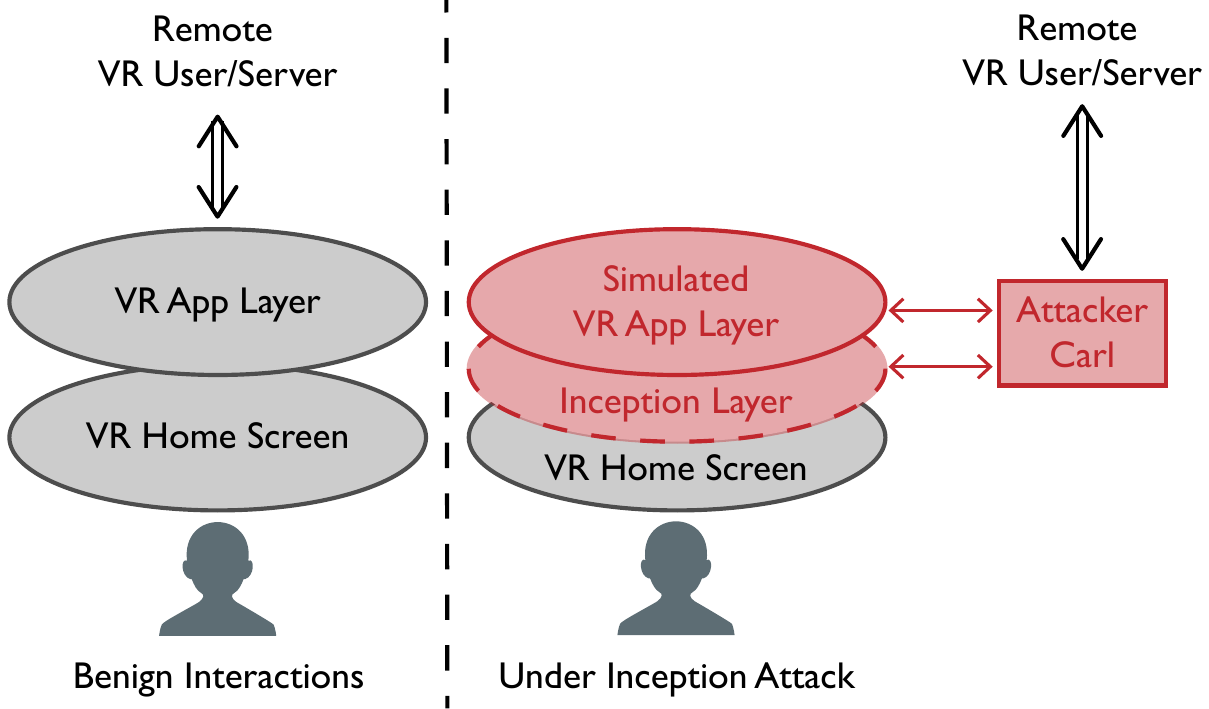}
\vspace{-0.1in}
    \caption{Inception Attacks: A user thinks they are interacting directly
      with a VR app launched from the VR home environment, when they are in fact
      interacting with a simulated environment inside the attacker's inception layer.}
    \vspace{-0.15in}
    \label{fig:intro}
\end{figure}

The flip side of such immersive interface is that when misused, VR systems
can facilitate security attacks with far more severe consequences~\cite{VRharassment} than
traditional attacks. By design, VR's immersive sensory channels create
a strong sense of realism and comfort for
users. However, attackers can exploit this trust to  
manipulate the immersive experiences, causing users to unquestioningly
believe and respond to what they see and hear inside VR,  whether it's
scenes, movements, or conversations. For example, consider the following scenarios:

\para{Scenario 1.} {\em
Alice spent the past two days taking her required
employee training through her employer's VR app. Today, she attempts the
final test using the same app. But despite her hard work, she keeps
failing. Frustrated and discouraged, Alice eventually gives up and
starts to doubt whether she is a good fit for the job. Elsewhere, her coworker Carl closes his
laptop with satisfaction.}
  
\para{Scenario 2.} {\em Alice takes off her headset in
  frustration, after the latest argument with her partner Madison during their
  regular VR date night. Madison had been increasingly distant all week and
  actually ended their relationship tonight. Elsewhere, Carl
  turns off his generative voice interface and watches as a confused
  Madison storms off in VR.}

\vspace{3pt} Both scenarios described above result from the same attack, where an attacker
(``Carl'') compromises the integrity of the user's VR system, and
inserts an attack software between the user and their VR interface. Under this
attack, users (``Alice'') are no longer interacting directly with
their expected VR counterparts (e.g., ``CompanyX's server,
Madison''). Instead, they operated on (and stayed trapped in) the attacker's VR
app, interacting with their VR counterparts through a hidden layer of
indirection.  These interactions are being monitored, recorded, and
modified in real-time by Carl.  
Because VR interfaces prioritize full immersion\footnote{In VR, each user is fully immersed
  in a single, highly engaging 3D environment that surrounds them on all
  sides, making them feel as if they are physically situated in
  the target space.}, there is no indicator for a user to understand who or
what is controlling their VR experience. 
We call this the {\em
inception attack}\footnote{The inception attack is inspired by both its
namesake, the 2010 Christopher Nolan film {\em Inception}, and movies with
similar attacks, including the 2018 Steven Spielberg film {\em Ready Player
One}.} and provide a high-level illustration in 
Figure~\ref{fig:intro}. 

Inception attacks are very powerful because of two properties. First,
an inception attacker can take near total control of the user's VR experience.
This includes silently observing a user's actions, recording input,
passwords, or other sensitive data, but also hijacking and replacing entire
social interactions in real-time.  In the aforementioned Scenario 2, Alice is
a target of an inception attack, and her VR date night with Madison
is hijacked. Now both Alice and Madison only see/hear what the attacker wants
them to see/hear. This is made even stronger with the addition of generative
AI tools capable of replicating human voices and generating deepfake
videos.

Second, inception attacks can be
extremely difficult to detect, because VR apps are often designed to
prioritize a realistic representation of the real world. Thus VR systems avoid 
using service indicators and security icons
typically found on traditional mobile or desktop computing
environments. 
Currently,  VR users have no reliable means of
verifying/authenticating which app is responsible for any aspect of their immersive
experience (visual or auditory).

In the rest of the paper, we demonstrate that inception attacks are viable
and extremely effective on current VR platforms today. We
describe our implemention of an inception attack on Meta Quest Pro headsets,
capable of hijacking the user's 3D home environment and their interactions
with 2D/3D apps (e.g., 2D Quest Browser and 3D VRChat).  We show how
attackers can alter a variety of user inputs and immersive interactions,
including financial transactions and live audio conversations in VRChat. We
include screenshots and video clips to illustrate these attack instances,
highlighting their feasibility and capability to produce harm.

We evaluate our attack instances via IRB-approved user studies, covering the 
entire attack process from seamlessly 
transitioning a user into
the VR inception app to hijacking their immersive interactions
with servers/users.  In our primary user study, our attack successfully deceived all
(but one) of 27 participants, including highly experienced users, who
use VR devices everyday. Even when we informed them of the
attack, half of our participants were unable to identify anything
unusual from their experience. The rest recalled very minor glitches as the session was
hijacked, yet all attributed these glitches to standard VR performance issues. 
Notably,  nearly all of the identified glitches were cosmetic
(e.g., cursor size) and
can be reduced or eliminated by fine-tuning the inception app and content.
Finally, we analyze  a range of potential defenses and
discuss their efficacy and usability tradeoffs.

Overall, our work makes {\bf three key contributions}:  
\begin{packed_itemize}\vspace{-0.02in}
\item To the best of our knowledge, we are the first to
  identify, implement and evaluate immersive hijacking attacks in
  VR.  We show that an inception attacker's capabilities extend 
  beyond eavesdropping on sensitive data~\cite{Munilla_Garrido_2024}; they can now
  manipulate and alter the immersive social interactions in VR, which are
  designed to replace physical interactions and are among VR's most
  compelling features.

\item  Our attack implementation employs a different methodology compared to 
  previous GUI confusion attacks on web browsers and
  smartphones~\cite{niemietz2011, Huang2012,niemietz2012, Bianchi2015, Fratantonio2017}.
Due to the full immersiveness delivered by 3D apps and their intensive
rendering requirements, VR naturally prevents any 3D app from injecting
GUI overlays and discreetly bringing itself to the foreground,  which are the basis of existing
  attacks.  Instead, our 
  attack leverages VR's natural operating process to transition 
  users into the inception app and trap them there undetected,  where the
  inception app contains a replicated 3D home
  environment and simulated ``apps'' launched from this fake home. 
  We
  also develop efficient methods to replicate/simulate those 3D 
  environments and apps.

\item We show that, as a maturing technology, current VR systems lack
  some of the
  security mechanisms deployed on traditional devices. Yet,
  implementing these mechanisms (or new ones) in VR is more
  challenging due to its immersive nature. We propose a multifaceted 
  defense pipeline to raise attack cost and reduce its likelihood
  of success.
\vspace{-0.07in}
\end{packed_itemize}
\para{Ethics.}  Our user studies were reviewed and approved by our
institute's IRB office. We disclosed both the attacks and potential
defenses to Meta, who responded with positive acknowledgment of our
disclosure. 

\secspace \secspace
\section{Background and Related Work}
\secspace
\label{sec:back}
In this section, we summarize existing attacks against VR systems and
GUI confusion attacks targeting web browsers and mobile devices.

\secspace
\subsection{Existing Attacks against VR}\label{subsec:existattackinvr}
\secspace
\para{Malware.}  Like all other computing
devices~\cite{mw_mobile_sec},  VR devices are naturally susceptible to
malware. Attackers can introduce malware directly onto a VR headset as 
malicious apps, or infect other devices connected to the headset. 
In fact, many VR privacy attacks 
assume a successful injection of malware to collect user 
motion data~\cite{vr_malware, HoloLogger, slocum}.

\para{Privacy Attacks.} To create immersive experiences, VR headsets
utilize an array of sensors to continuously monitor user activities,
which may lead to privacy attacks. Numerous studies show that
records of a user's head and body motions can expose sensitive
information, including identity, age, gender, ethnicity, emotions,
health conditions, and even the video content they have watched~\cite{Nair_2023,
  hidden_reality, Munilla_Garrido_2024, sabra2023exploiting,
  deanonymization, identification50000, 360videousenix24}. In parallel, researchers have exploited ways to
eavesdrop on specific apps or user input. The man-in-the-room
attack~\cite{mitr} infiltrated the Bigscreen VR app to allow unauthorized
users to invisibly join and observe their private chat rooms.  A range of
keystroke inference attacks can recover typed content by either observing avatar hand movements~\cite{yang2023virtual},
reverse-engineering an app's network packets~\cite{ucsbkeystrokeusenix24}, 
or analyzing side-channel leakage patterns~\cite{side_channel, AcousticKeystroke}. 

\para{Perceptual Manipulation.}  These attacks seek to manipulate user
orientation/movement in a virtual session by changing their perceived environment. User studies have shown
that, if these attacks can be implemented in practice, they can deceive
users into changing physical locations, cause 
physical collision, or induce motion sickness~\cite{manipulation,
  human_joystick}.  A straightforward implementation is to develop a malicious app
that shifts users virtual locations when they use the
app~\cite{Tseng_2022}.  Others assume the attacker has strong
device privileges to modify the screen feed of mixed reality (MR) 
headsets~\cite{manipulation} or the system
configuration files of a VR headset~\cite{human_joystick}.   Finally,
recent works also studied UI security in augmented reality
(AR) because multiple processes/apps can place customized content on a
AR device's screen~\cite{uipropertyAR,sharedstateAR}.  

\secspace
\subsection{GUI Confusion Attacks}
\secspace
The inception attack is related to a broader class of GUI attacks,
with analogous attacks in the context of web browsers~\cite{Huang2012, Ryck2013}
and mobile devices~\cite{felt2011, Bove2022}.  We summarize these GUI attacks
and highlight their key differences with the VR context.

\para{Browser-based GUI Attacks.}  The closest analogy to inception
attacks 
in web browsers is {\em clickjacking} attacks~\cite{niemietz2011,
  Huang2012}. By abusing HTML/CSS features, a clickjacking attack overlays a
transparent frame onto a legitimate page, deceiving users into clicking on
unintended UI components. Modern web browsers prevent these 
attacks by implementing frame busters~\cite{niemietz2012} (to detect and
extract frame overlays) and additional security measures like Content
Security Policy~\cite{OWASP}.

\para{Mobile GUI Attacks.} There are similar GUI-based overlay attacks on
mobile operating systems such as Android. Attackers can either place
customized overlays to obscure individual UI components or the entire UI
window~\cite{niemietz2012, Bianchi2015}, or exploit the Android developer API to obtain
partial control over the system task stack and elevate a background process
to replace a foreground app's UI~\cite{Bianchi2015}. These attacks can enable
clickjacking, allowing attackers to steal a user's login credentials or
install apps/permissions without user's awareness~\cite{Fratantonio2017}, and
are further facilitated by side-channel attacks that monitor UI states in
real-time~\cite{Chen2014}.

A number of techniques can help mitigate these attacks by verifying the
authenticity of the GUI.  These include the Android Window Integrity (AWI)
model that tracks foreground UI states and flags suspicious
activities~\cite{ren2017}, an app vetting tool using static
analysis~\cite{Bianchi2015}, and displaying trust information of the
foreground app and a secret image on the navigation bar~\cite{Bianchi2015}.
In addition, updates in the Android APIs allow developers to
specify UI elements to be protected (where OS ignores inputs if another UI window
overlaps them), and overlays can no longer obscure critical UI components
such as the status bar, navigation bar, and confirmation
dialogs~\cite{Bove2022}.

\para{Distinction from VR and Our Work.} VR inception attacks are similar in
intent as these GUI-based attacks, but are quite distinctive given the constraints
of the VR platforms. First, traditional GUI attacks cannot function in VR,
because the attacking app needs to either embed customized overlays on
the GUI or covertly switch itself to run in the foreground. In contrast, VR
systems today only display a single 3D environment, and does not support
overlays. And since background apps cannot render its 3D environment in the background, an
attacker app jumping into the foreground would cause a very noticeable disruption
to the UI.  A fullscreen GUI attack on Android mobile devices has been
hypothesized~\cite{Bianchi2015} but not designed or implemented.

Another key distinction between VR platforms and traditional operation
systems is that inception-based attacks will be more challenging to
secure in VR
platforms compared to prior GUI-based attacks. First, VR platforms lack the broad
range of mechanisms that authenticate endpoints and information flow
available in desktop, mobile and browser-based operating systems. For
example, there are no abstractions that authenticate VR connections to VR
providers, analogous to SSL connections, CAs and server certificates. Second,
VR systems are designed to be fully immersive, a goal that often conflicts
with requirements of visible security indicators. Navigating these tradeoffs
to design usable and effective security indicators in VR is an active and open
challenge.

\secspace
\section{Inception Attacks}
\label{sec:inception}
\secspace

In the context of VR, we define an {\bf Inception
  Attack} as:
\vspace{2pt}
\fbox{\begin{minipage}{23em}
\emph{An attack where the attacker controls and manipulates the
  user's interaction with their VR system,  by trapping the user inside
  a single, malicious VR app that masquerades as the full VR system.}
\end{minipage}}
\vspace{2pt}

\noindent An inception attack inserts an attack layer between the user and any
external entities, behaving as a hidden layer of indirection. 
While the user thinks they are interacting normally with their home
environment,  apps and other users,  they are interacting
with a {\em single} malicious app, where everything they input, see and hear
can be intercepted, relayed, and possibly altered by the
attacker. These include both user input (e.g., voice, motion,
keystrokes)  and app
output (e.g. content from app servers, actions of other users).
Therefore, the inception attack has a significantly broader
scope and higher potential for damage than existing VR 
attacks (\S\ref{subsec:existattackinvr}).

At a high level, the inception attack is analogous to a
man-in-the-middle (MITM) attack in a network,  but applied to the VR
context. Similar to MITM attacks, inception
attacks can be executed by exploiting various security
vulnerabilities in today's VR systems. The specific attack implementation may
vary depending on the attacker's access privileges within the target
VR headset.  In the following, we discuss the threat model considered
by our work and provide an overview of the attack design.

\secspace
\subsection{Threat Model} 
\secspace
\label{sec:threat}

\para{The Target.} We consider a target user who has a standard, clean VR  
headset with developer mode activated. 

There are three reasons why this is quite reasonable. First, developer mode
is commonly supported today by a number of VR platforms (Meta Quest, VIVE
Focus and PICO), both to improve usability and to unlock appealing features
such as adjusting headset resolution, capturing screen content and debugging.
Second, there are already reports~\cite{sidequest} of millions of VR users
today actively enable developer mode for expanded features and
because it is required by popular enterprise-grade VR management solutions
like ArborXR~\cite{arborXR}.  Finally, for Andoid devices, developer mode is
a persistent setting that, once activated, remains enabled until manually
deactivated. 

\para{The Adversary.}  We make two key assumptions on the adversary.  First,
we assume the adversary does not have physical access or root access to the
target's VR headset, but can leverage developer mode to run apps and scripts
on the target's VR headset (we discuss multiple ways to do so in
\S\ref{subsec:bootstrapping}).  Second, we assume the adversary has
reasonable computing resources to build VR apps, an external server to
communicate with the attack app, and their own standard VR headset (same
model as the target's).

\para{Alternative Threat Models.}  We consider the above threat model because
it is realistic and is the most likely set of conditions for the
attack. However, it is not the only possible threat model.  Later in
\S\ref{sec:appattack}, we broaden our discussion to include alternative 
threat models that also enable the inception attack.

\secspace
\subsection{Overview of the Attack Process}
\label{subsec:designoverview}
\secspace
The attacker performs three sequential tasks to achieve the goal of hijacking the
target's interaction with their VR system.

\begin{figure}[t!]
    \centering
      \includegraphics[width=0.8\linewidth]{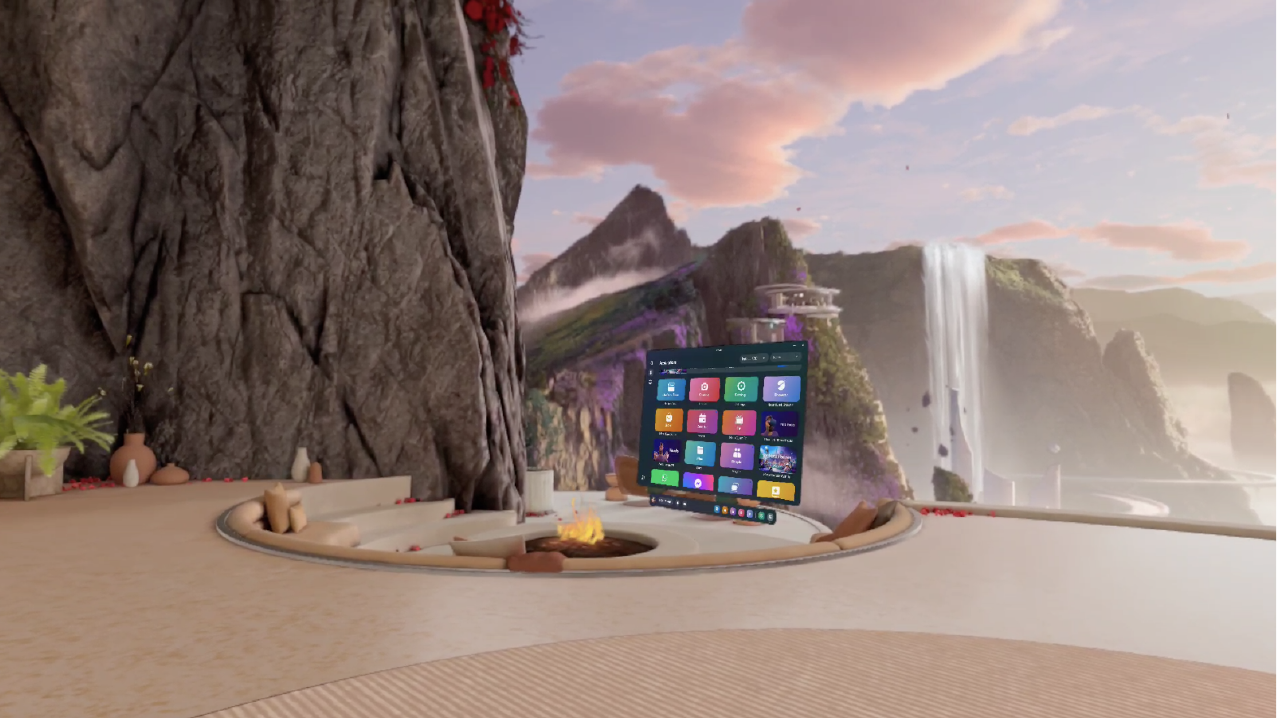}  
      \vspace{-0.05in}
      \caption{Screenshot of an example 3D home environment with a menu panel for apps, captured on a Meta Quest Pro headset.} 
      \label{fig:home_env}
       \vspace{-0.1in}
     \end{figure}

\para{(1) Build the inception app.}  The attacker collects headset
configuration information and uses it to compile an inception app customized
for the target. Hidden in this version is the replicated 3D home environment
(see Figure~\ref{fig:home_env}) and simulated versions of the apps installed
on the target headset. Note that this is a one-time cost. Once compiled,
  an app can be easily customized to target different users and their configurations.

\para{(2) Activate the inception attack.} The attacker installs the
inception app on the target's headset, and runs a background script on
the headset to detect when the target signals the system to 
exit a currently active app. Normally, this signal would prompt
the system to return the user to the benign home environment. But the attacker script intercepts and destroys this
signal, terminates the current app and initiates the inception app. This
silently transitions the user into the simulated home environment, 
and starts the inception process.

Our decision to activate inception during an app exit is deliberate, as each
app exit creates a natural disruption in the user's immersive experience.
The attack can begin seamlessly while avoid user suspicion.

\para{(3) Hijack user interactions.}  Once the headset enters the
attacker-replicated home environment, all interactions between an user and the VR system
will go through the app, including the apps the user sees and uses.  Thus the
attacker can control and modify user interactions with any ``simulated''
app displayed on the fake home environment.  Using the method
described in (2), our attack ensures that exiting 
any app will simply return the user to the fake home environment. As
such, the user is trapped inside the inception app. 

\secspace
\section{Attack Implemention on Meta Quest VR}
\label{sec:attackdesign}
\secspace
We now present an implementation of the inception attack on Meta Quest  headsets, applicable to all three device versions (Quest 2/3/Pro).   In this section, we detail the implementation of the first two tasks in the attack pipeline.  Later in 
\S\ref{sec:cases}, we describe the third module using two
specific apps: Quest Browser (2D) 
and VRChat (3D).

To provide context, we first describe the workflow and features of Meta Quest headsets, and how the adversary gets to install and run apps and scripts on the target's Quest headset. 

\secspace
\subsection{Preliminaries: Meta Quest VR}
\label{subsec:prelim}
\label{subsec:bootstrapping}
\secspace
Meta Quest headsets run on Horizon OS~\cite{horizonOS}, a modified version of Android OS. After putting on the headset, a user first sees their
3D home environment and a menu panel of apps (Figure
\ref{fig:home_env}). The user clicks on an app icon from the panel to
enter the app, and returns to the home environment after exiting the app, typically by pressing the ``home'' button on the right controller.

\para{Developer Mode. }  As outlined in our threat model, we assume the
target has activated the developer mode on the VR headset. This is 
done by toggling a switch in the Meta Quest app on the user's smartphone.
Note that developer mode is not perceived as a sensitive configuration like root access, 
as it neither grants root access to the headset nor permits any root-level
operations.

\para{Android Debug Bridge (ADB).} Enabling developer mode {\em
  automatically} activates the Android Debug 
Bridge (ADB), making the headset ready to accept ADB commands. ADB is
a command-line tool that allows an individual to command an Android
device (e.g. a Quest headset) with flexibility. It  supports functionality
like running shell scripts, installing apps, reading device settings, and transfering data via wireless.  Disabling ADB requires deactivating developer mode.

Via ADB, a machine can command the device by requesting USB or wireless connection to the device.  Furthermore, after installation, an app with ADB client and server packed inside its library, can request access behaving as a wireless ADB connection~\cite{shizuku_usermanual,shizuku}. Today, it is quite common for benign apps and peripherals to have (wireless) ADB access to Android devices~\cite{ocularmigrain, automate, shizuku}.  Many of these apps (e.g.~\cite{automate}) have millions of active users. 

\para{Attacker Obtaining Wireless ADB Access.}  An attacker machine can run scripts to command a headset remotely by requesting wireless ADB access to the headset.  Obtaining  such access only requires the machine to be on the same WiFi network as the headset, and requires no credential-based
authentication~\cite{rsakey}.   Since it is common for apps/peripherals to require wireless ADB access, the user would 
  perceive this request as a harmless request (from installed
  apps or peripherals\footnote{All the wireless ADB requests only display a ``Allow
    USB debugging?'' pop-up and the requesting device's RSA fingerprint. This information is insufficient for users to identify the device type and owner, as shown by~\cite{rsakey}}.)  and grant 
  access by default.  Now the attacker can run shell scripts and install apps on the
target headset without the user's awareness.

\para{Attacker Publishing an App with ADB Embedded.} Alternatively, 
the attacker compiles an app (i.e., the inception app) with ADB client and server packed
inside. The attacker disguises this app as a useful
utility app,  publishes and promotes it on popular app sites like
SideQuest~\cite{sidequest}.  When the user installs and runs this
app, they must permit the app with ADB
access for it to function.  Again, access will be easily granted given the
frequency of benign ADB requests.
Now the app can silently run shell scripts in the
background and communicate with the attack server.  The scripts 
continue to run even after the user exits the app. 

Either approach can facilitate the attack, whether the target is a particular
user (first approach) or an opportunistic attack seeking to spread the attack
broadly (second approach).
Since the attack proceeds nearly identically for both approaches, we will
proceed our discussion assuming the attacker uses the first method.  

\secspace
\subsection{Constructing the Inception App} 
\label{subsec:buildapp}
\secspace
Leveraging ADB access,  the attacker runs a shell script to quickly collect headset configuration, and uses them to compile an inception app customized for the target. This customized app contains a replicated 3D home environment used by the target user and simulated versions of the apps installed on the target headset.  

\secspace 
\subsubsection{Collecting Headset Configurations}
\secspace
\label{subsec:configuration}
The attacker needs two configuration data: the 3D background used by the home environment, and the list of installed apps and the state.  Since 3D backgrounds are stored as APK files on the headset, the attacker only needs to identify the APK file used by the current home.   The ADB command \verb|adb logcat|  returns a log of system activities including the background APK in use. Since 3D backgrounds typically come from Meta Quest's built-in options or SideQuest Custom Home~\cite{customhomes}, once identified, the attacker can obtain the APK from their own headset or download it from SideQuest.

The command \verb|adb shell cmd package list packages| returns a list of installed apps on the headset. Knowing the app names, the attacker can obtain their button images, from either app APKs on their own headset or online sources like Meta App Store. Finally, the attacker can access the state information, such as recently used apps, using \verb|dumpsys|.

\secspace
\subsubsection{Replicating the 3D Home Environment}
\label{subsec:rephome}
\secspace
One can accurately replicate the target's 3D home environment by focusing on three key aspects: immersive content (the 3D background and app menu, plus monitoring/responding to user inputs), managing app launches and exits, and displaying/configuring device settings.  

Replicating the immersive content is straightforward: unpack the background APK file to retrieve the 3D background, arrange app button images in a grid, and display state information, like the most recently opened apps.  Other static configurations (cursors, pointers, and the menu bar) can be replicated with high precision. Similarly, recognizing and responding to user inputs is streamlined with Meta's Interaction SDK~\cite{metasdk}, which allows developers (including the attacker) to  integrate user interactions into their app.

\begin{figure}[t!]
  \centering
    \includegraphics[width=0.75\linewidth]{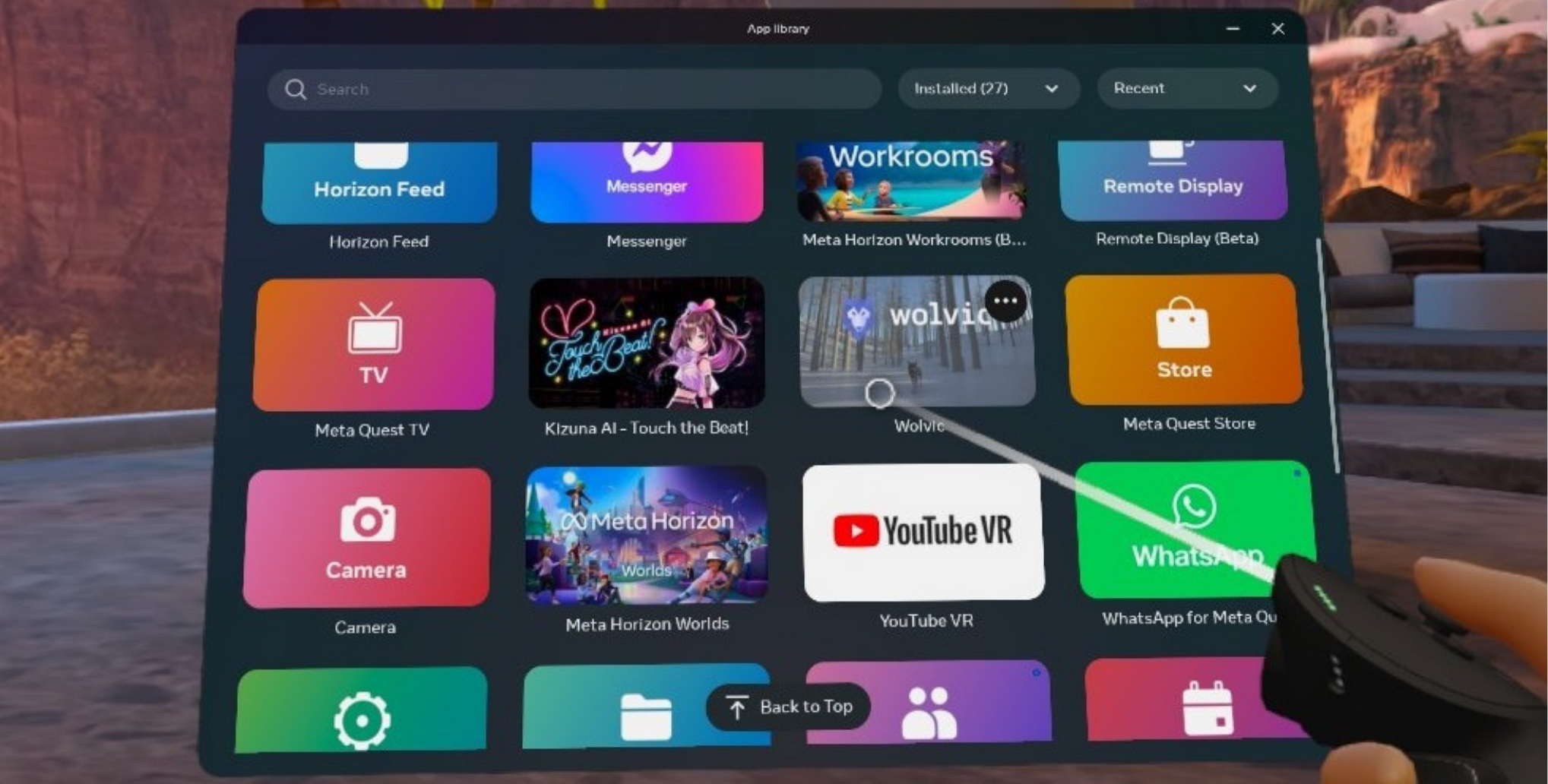}  
    \vspace{-0.05in}
    \caption{The app library window in the Meta Quest Pro VR home. The user is selecting an app called wolvic.} 
    \label{fig:app_lib}
    \vspace{-0.1in}
\end{figure}

When the user clicks on app A's button on the home menu (Figure~\ref{fig:app_lib}), the inception app will transition into running a simulated version of A.  When the user exits A (by pressing the controller's home button or the virtual exit button), the attack script running in the background intercepts the exit signal, halts any subsequent activities, and returns the user to the replicated home environment in the inception app.

Finally, since full immersion requires access to key device elements include WiFi~\cite{androidnetwork}, Bluetooth~\cite{metakeyboard} and audio~\cite{metaaudio}, they are automatically granted to all apps.   Thus it is easy to display/change them as device settings in the replicated home.

\para{Removing App Splash Screen.} One notable distinction between an app and the native home environment is that an app often displays splash screen at launch time (e.g. all apps created with Unity include one by default). This is not the case for the home environment. When packaging the inception app, we easily eliminate the splash screen by using the free Unity Asset Bundle Extractor to adjust app parameters~\cite{removesplash}. 

\secspace
\subsubsection{Simulating the Installed Apps}
\label{subsec:repapp}
\secspace

Simulating all installed apps might initially seem overwhelming, as replicating the development efforts for each app is impractical. Instead, we show that simulating individual apps can be efficiently accomplished through four distinct methods, each providing varying degrees of hijacking capability.

\para{Direct App Call (no hijacking).}  This is for apps that the attacker does not wish to hijack/control, but needs to replicate to avoid suspicion.   Here, the attacker only needs to instruct the fake home environment to call the actual app (installed on the headset).  The action of an app sending the user to another app is already supported by Android as one of its main features~\cite{androidintents} with ready-to-use plugins~\cite{applauncher}.  When the user exits this app, the attacker can again transition the user back to the fake home (details in \S\ref{subsec:launch_inception}).

\para{Replicating App GUI (stealing credentials).}  If the attacker only wants to eavesdrop on the target's credentials on an app, they only need to replicate the login page of the app. After the target inputs their credentials on the replica, the attacker script loads the original app to the foreground.  Notably, if the original app includes a splash screen during loading, this will lead to an extra splash screen that may raise suspicion.  One way to mitigate is to first display an app crashing popup and then load the app. Given the common occurrence of Meta Quest apps crashing during loading, this helps minimize suspicion.  Finally, upon obtaining the credentials, the replica switches to the dormant mode (i.e. directly loading the original app) until there is a need to record credentials again. 

\para{Local Simulation via API Calls (full hijacking).} This aims to realize the full immersive hijacking on the app. Although the attacker generally lacks access to the source code, we show that simulation can be efficiently achieved by replicating the app's GUI and API calls. This is because many VR apps opt to use public APIs~\cite{publicapis} to streamline app development and expand compatibility across VR platforms~\cite{api_app}.  They use these APIs to interact with app servers and fetch data to display on the headset. 

In this case, the replica first clones the app's GUI and user interaction, similar to the process of replicating the 3D home (\S\ref{subsec:rephome}).  It calls the public APIs to communicate with the app server to fetch data, using the target's credential obtained via the replicated GUI.  After receiving server data, the replica updates the GUI to display content on the headset.   The end result is that the attacker can hijack the target's app session in  multiple ways:  (1) eavesdrop on communications between the target and the app server; (2) change the target's input to the app;  (3) modify app data displayed on the target's headset.  Later in \S\ref{subsec:browserimplt} we use this method to simulate and hijack the Meta Quest Browser app.

\para{Over-the-network Simulation (full hijacking).} For all other apps,  we design an over-the-network simulation, where the attacker uses a separate VR headset to run the target app.  Now the app replica includes two coordinating components:  one on the target's headset and another on the attacker's headset connected to a computer via USB.  The app replica on the target's headset implements the app GUI and user interaction (like the above),  from which it obtains the target's credentials for the app and forwards them to the attacker's computer using standard network connections.  Using the target's credentials, the attacker's headset runs the original app to interact with the app server. As such, the attacker's headset/computer combination acts like a MITM between the target and the app server,  allowing the attacker to manipulate the interaction between the two.  Later we describe the details of over-the-network simulation of the 3D VRChat app in \S\ref{subsec:vrchat}.

\secspace
\subsection{Activating the Inception}
\label{subsec:launch_inception}
\secspace
After packaging an customized inception app for the target, the attacker installs it on the target headset via ADB.  In our experiments, this app is around 700MB, primarily
  due to the 3D home scene used by the target. This size is
  comparable to most VR apps and well within the official 4GB 
  limit. 

  The next step is to start the inception session. As outlined in \S\ref{subsec:designoverview}, the optimal moment to covertly transition the user into the inception is upon them exiting an 3D app, where they expect to return to the home environment. Instead, the attack brings them back to the replicated home environment without raising suspicion.   The attacker achieves this by running a shell script on the headset (see Algorithm \ref{code:spy} in Appendix).  It runs in the background to monitor  the headset activity using ADB commands \verb|getevent| and \verb|dumpsys|. Upon detecting any signal of an app exit, the script proceeds to stop/remove  subsequent activities and activates the inception app. The script continues running and performs the same action whenever it detects an app exit signal, effectively trapping the user inside the inception app.

\secspace
\section{Detailed Hijacking Attacks}
\label{sec:cases}
\secspace 
To demonstrate the feasibility and impact of immersive hijacking, we simulate two representative VR apps in our inception app and use them to perform multiple hijacking attacks. The two apps are Meta Quest Browser~\cite{questbrowser}, the built-in web browser for Quest headsets, and 3D VRChat~\cite{vrchat}, a social interaction app with more than 20 million monthly users.

\secspace

\subsection{Hijacking Meta Quest Browser}
\label{subsec:browserimplt}
\secspace
As the built-in browser for Quest headsets,  Meta Quest Browser offers immersive web browsing experiences.  Users can navigate through interactive websites, enjoy streaming videos on a captivating theater-sized screen, and engage with social media platforms seamlessly.  Its popularlity has motivated web developers to build attractive, novel web content with immersive experiences.

\begin{figure}[t!]
  \centering
    \includegraphics[width=0.99\linewidth]{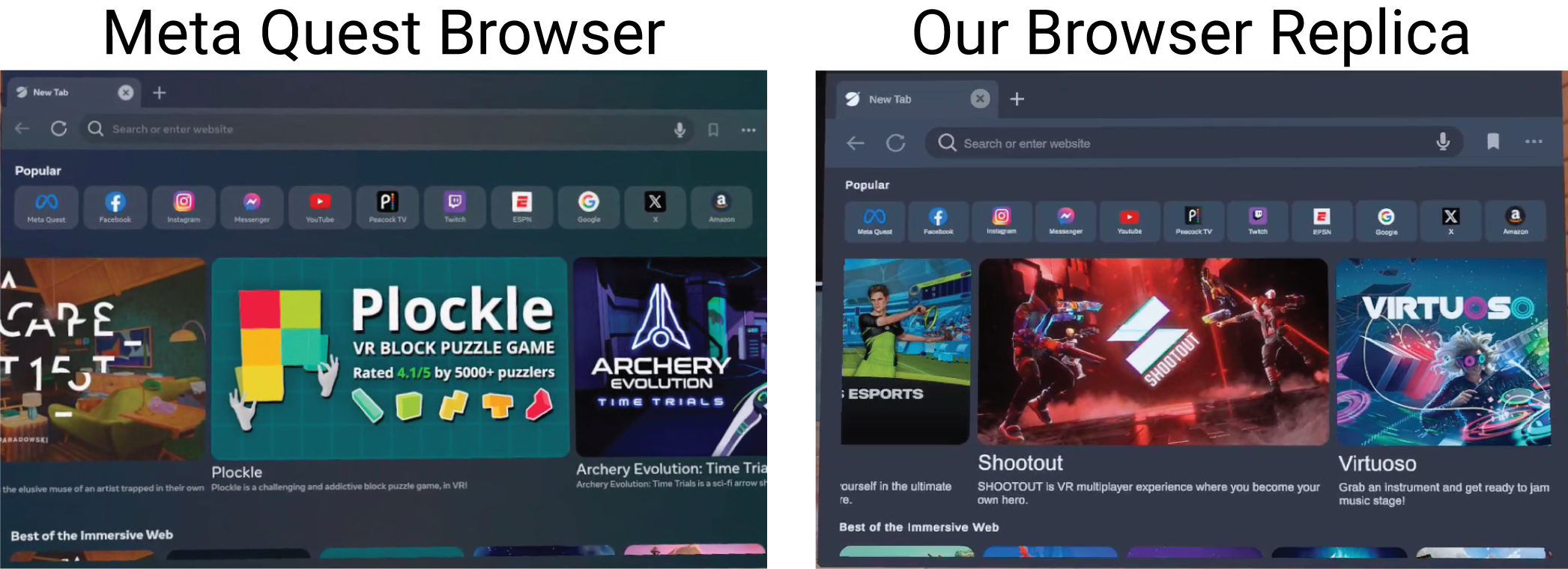}  
    \vspace{-0.05in}
    \caption{A side-by-side comparison of Meta Quest Browser and our replica.  Full size images at Appendix,  Figure~\ref{fig:browser_fullsize}.} 
    \label{fig:browser}
    \vspace{-0.05in}
\end{figure}

\para{App Simulation.} We build a simulation of the browser app in Unity~\cite{unity} and package it into our inception app. Since  Quest Browser uses public APIs to communicate with web servers (e.g. HTTP requests to access websites),  our effort mainly focuses on replicating the browser GUI, displaying website content, and detecting and responding to user inputs.

We clone the browser GUI using Unity UI~\cite{unityui}, creating a canvas object as the browser window. We configure it to display various UI elements, including texts, images, buttons, toggles, webpages and more. Unity UI also handles registration of user interactions. For instance, each button element has a \verb|OnClick| function to define the effect of clicking it.  To display web content, we use Android WebView library~\cite{androidwebview,vulpex} (\verb|loadUrl(the_URL)|) to  fetch the content of a website at \verb|the_URL|, and display the content at the corresponding location on the canvas.  To access user inputs, we use Meta's Interaction SDK~\cite{metasdk} to monitor and log user's clicks, drags, and keystrokes on each webpage and their exact locations, and respond to these actions by updating the GUI content and/or sending API calls to web servers to fetch desired contents.  Figure~\ref{fig:browser} provides a side-by-side comparison of screenshots from the actual Meta Quest Browser and our replica. Our replica closely mirrors the real version. 

\para{Specific Hijacking Attacks.}  Using this replica, we successfully executed three attack instances: eavesdropping, modifying website content displayed on the headset, and modifying user input to the browser, all in real-time.  We describe each below, including screenshots of attack results. 

\vspace{4pt} {\bf Attack 1: Eavesdropping on sensitive data}: The target uses Quest Browser to access sensitive accounts in banking, corporate, and medical websites.  The replica can intercept and log private data, including credentials entered by the target.  This is because, using Meta interaction SDK, the replica can accurately track and record cursor movements, keystrokes, button presses, and headset motions.  It also has complete knowledge of the website's visual organization.  Thus, the attacker can extract user inputs to specific web entries.

\begin{figure}[t!]
    \centering
      \includegraphics[width=0.72\linewidth]{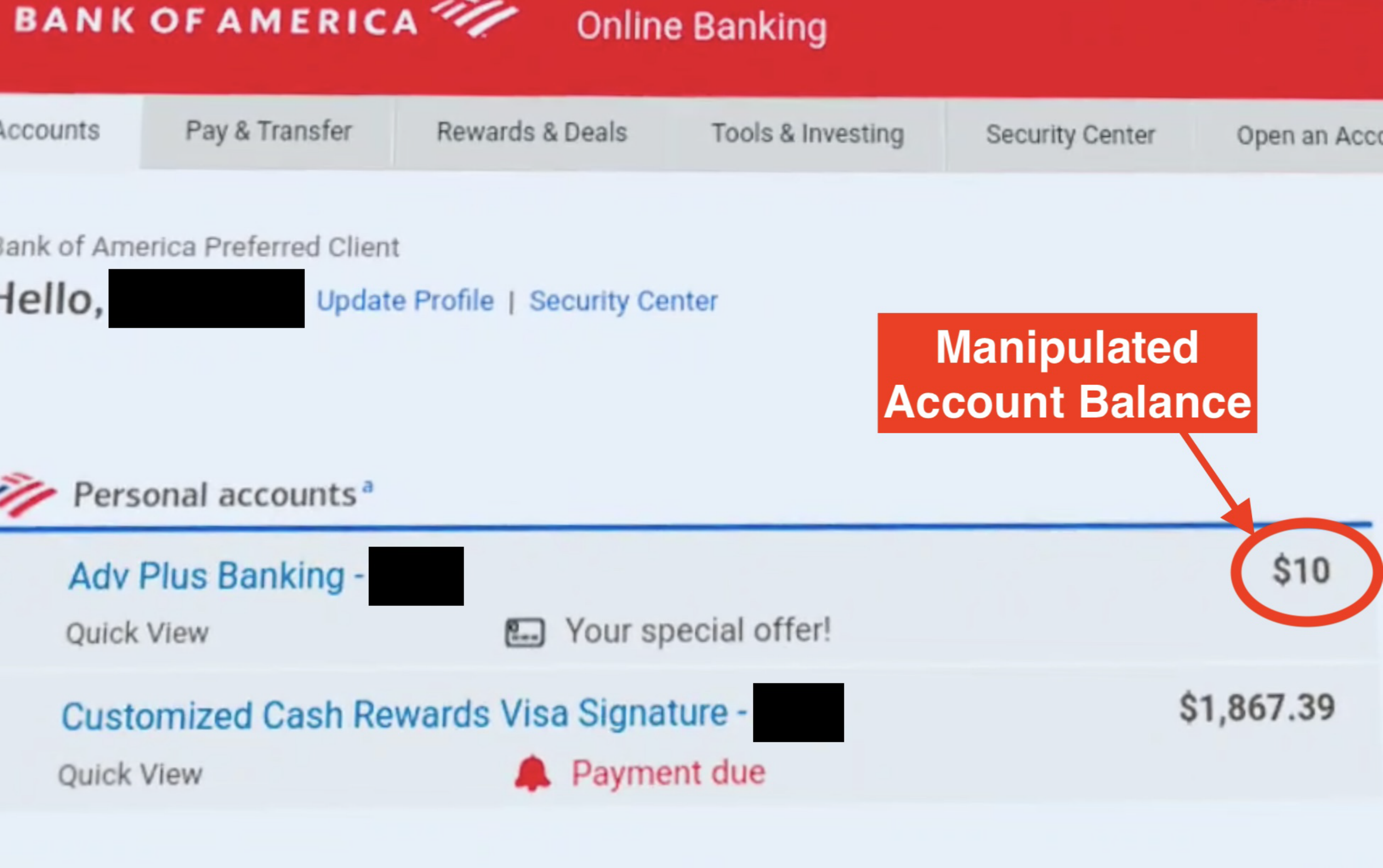}  
      \vspace{-0.050in}
      \caption{In a banking scenario, the Bank of America server sends the
        correct account balance to the headset. Our attack
        modifies this balance to display \$10 on the headset screen.} 
      \label{fig:bank_hack}
      \vspace{-0.1in}
\end{figure}

\vspace{4pt} {\bf  Attack 2:  Manipulating displayed content (e.g. bank balance)}: The replica browser can display a modified version of the website content fetched from the server.  Consider a banking session.  When the target uses the replica to access a bank website,  the replica first collects their credentials from its GUI and sends the credentials to the bank server via HTTP requests.  After verifying the login credentials, the bank server returns the target's account information to the headset, including the account balances.  While all of these network communications are encrypted using SSL handshakes,  the content to be displayed on the headset is encrypted using a key provided by the replica during the handshake. Thus the replica can decrypt and obtain the raw content, and can modify them (e.g. using JS code) before displaying them on the headset. 
Figure~\ref{fig:bank_hack} shows a screenshot of the target's headset display, where we altered their Bank of America account balance to $\$10$ using few lines of JS code (see Appendix~\ref{appendix:sec5jscode}). 

\begin{figure}[t!]
  \centering
    \includegraphics[width=0.99\linewidth]{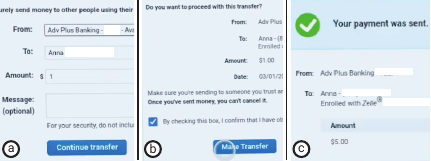}  
    \vspace{-0.05in}
    \caption{In a Zelle transaction, the transfer amount submitted by the
      target is altered by our attack before reaching the bank server. (a)
      The target inputs a \$1 transaction on the web
      form. Our attack secretly alters it to \$5 before sending it to
      the server. (b) The target is then taken to a confirmation page to
      finish the transaction, where our attack sets the amount to \$1 on this
      page to avoid user detection. (c) The actual transaction amount is \$5,
      confirmed by the bank. See Figure~\ref{fig:transaction_hack_fullsize}
      for a full-sized view.}  
    \label{fig:transaction_hack}
    \vspace{-0.1in}
\end{figure}

\vspace{4pt} {\bf Attack 3:  Manipulating target's input (e.g. payment amount)}: The attacker can also modify user inputs to the browser replica before using them to generate API calls to web servers. Note that these API calls typically use plain text and numerical values in the parameter fields. 

Figure~\ref{fig:transaction_hack} demonstrates a scenario where the target initiates an e-payment through Zelle in VR, a digital payment service owned by Bank of America. The target requests a \$1 transfer by completing the web form and clicking ``Continue Transfer'' (Figure~\ref{fig:transaction_hack}a). This takes the target to a confirmation page (Figure~\ref{fig:transaction_hack}b) to finish the transaction by clicking ``Make transfer''.   Using a few lines of JS code (Appendix~\ref{appendix:sec5jscode}), we successfully modify the transaction amount to \$5 and modify the amount displayed on the confirmation page to be \$1.  Thus the HTTP request sent to the bank server will request \$5 while the user perceives that \$1 is requested. Finally, Figure~\ref{fig:transaction_hack}c shows the legitimate bank confirmation where the actual transaction made is \$5.  Note that this page should also be modified by the attacker in a practical attack.  We leave it this way to demonstrate the success of the attack.

\secspace
\subsection{Hijacking VRChat}
\label{subsec:vrchat}
\secspace
In \S\ref{sec:intro} we described an inception scenario where Alice and Madison's VR date night session is  hijacked by Carl. We now describe an implementation of this attack on Quest devices by simulating the 3D VRChat app~\cite{vrchat}, where the attacker can listen to and modify live audio conversations between Alice and Madison.   Note that while the attacker hijacks the interaction of two users,  they only need to launch the inception app on one (i.e. Alice). 

We choose VRChat because it is the leading platform for VR users to interact with each other through personalized avatars, with more than 20 million monthly users~\cite{vrchatusers}. In its immersive 3D environments, participants engage in conversation, gaming, and romantic encounters like date nights~\cite{vrdating}. To safeguard player privacy, VRChat uses private APIs for communication between players and servers.

\begin{figure}[t!]
      \centering
        \includegraphics[width=0.6\linewidth]{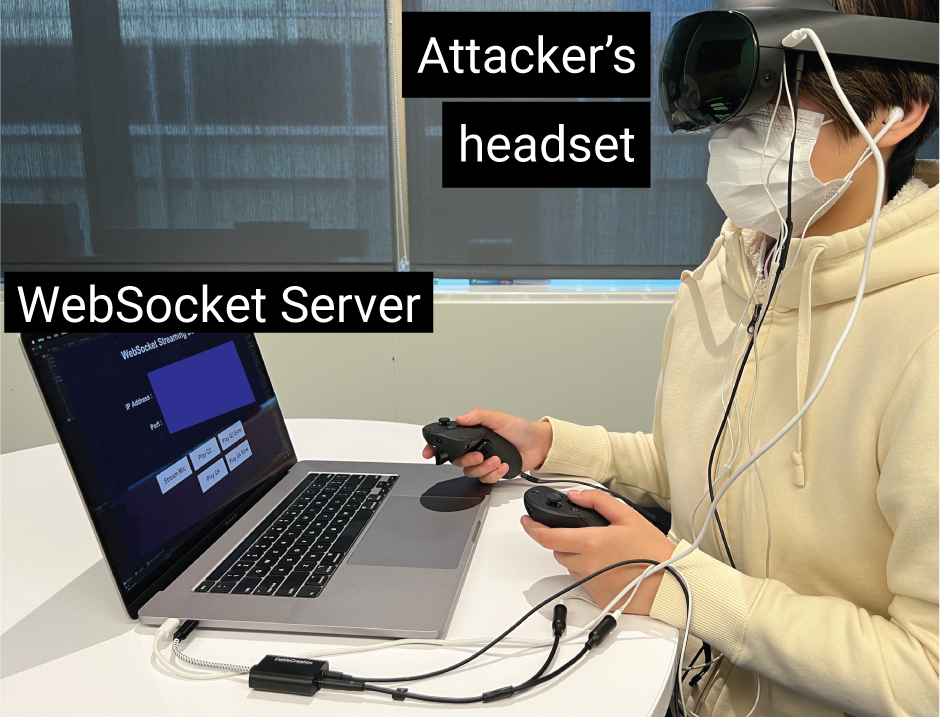} 
        \caption{A physical setup of the laptop/headset combo used by the attacker.  We run a WebSocket server on the laptop which receives Alice's livestream audio,  and feed it into the attacker headset by mounting the earbuds of a wired earphone near the headset microphones. Meanwhile, the audio output of the attacker headset (i.e. Madison's speech) is feed to the laptop via an audio cable and then livestreamed to the VRChat replica on Alice's headset.}
        \vspace{-0.10in}
        \label{fig:earphone}
  \end{figure}

\begin{figure}[t!]
    \centering
      \includegraphics[width=0.825\linewidth]{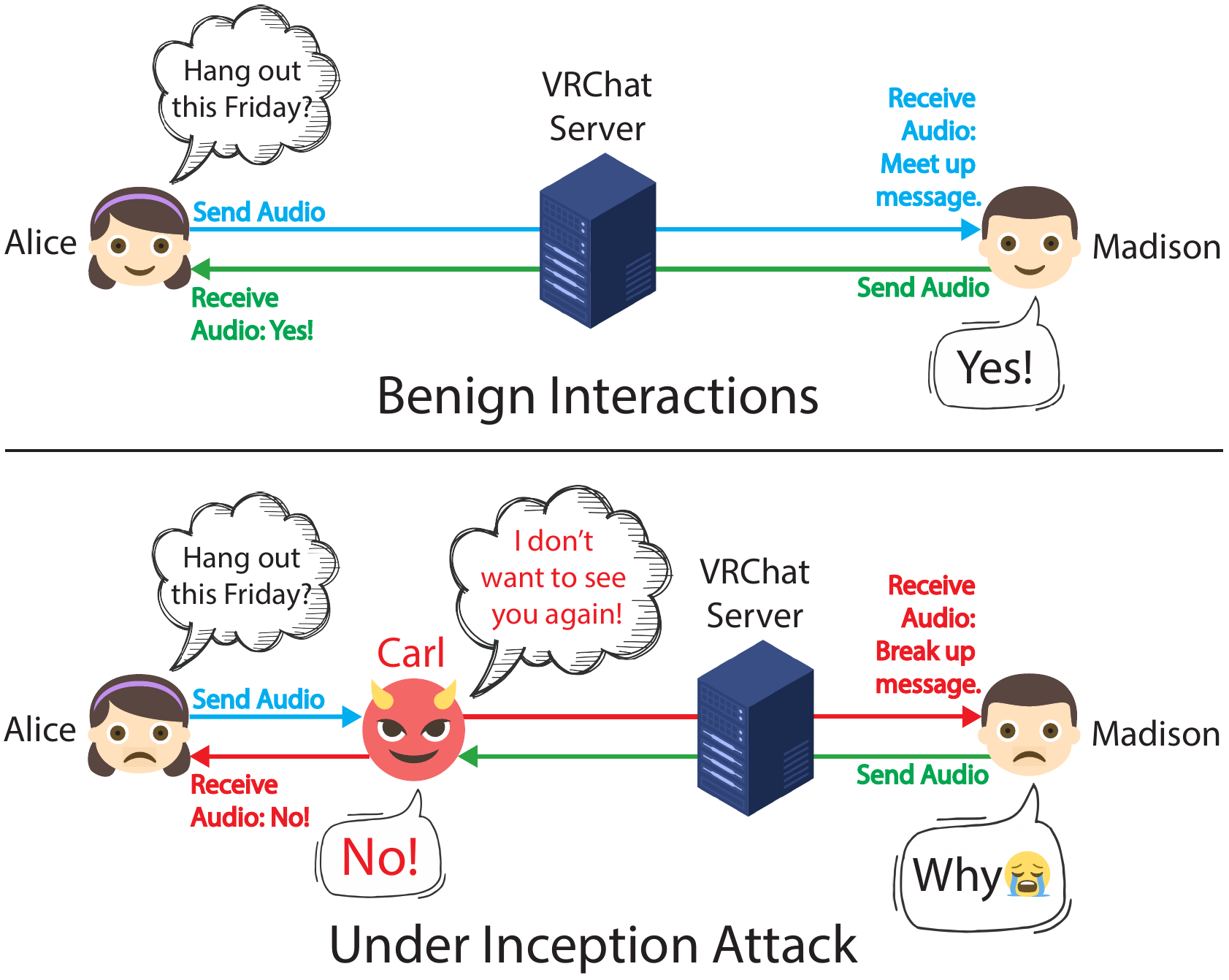}  
      \vspace{-0.10in}
      \caption{The attacker Carl hijacks Alice and Madison's VRChat session.  Carl monitors and modifies live audio communications, so Alice and Madison hear different conversations.  For example, Carl made Alice hear Madison answering ``No'' to her question of ``Hang out this Friday?'' although Madison responded ``Yes''; Carl crafted Alice's speech of ``I don't want to see you again!'' and sent it to Madison to enforce the breakup. }
           \vspace{-0.10in}
      \label{fig:vrchat_abstract}
\end{figure}

\para{Over-the-Network App Simulation.}  We replicate VRChat using over-the-network simulation (\S\ref{subsec:repapp}). We setup a laptop (Macbook Pro 2019) and an additional Meta Quest headset connected to the laptop via USB (see Figure~\ref{fig:earphone} where the attacker wears the headset). 
This combo acts as a MITM between the target and the VRChat server.  In the following,  let Alice be the target of the inception attack and Madison be the one interacting with Alice in VRChat.  Figure~\ref{fig:vrchat_abstract} sketches the attack flow where attacker Carl hosts the laptop/headset combo, in comparison to the benign flow.

To bootstrap the attack, the replica on  Alice's headset replicates the GUI of VRChat, from which it captures Alice's credentials (of the VRChat account),  using the ``stealing credential'' method discussed in \S\ref{subsec:repapp}.  This enables the attacker headset to run a legitimate VRChat app using Alice's credentials, which communicates with the VRChat server, and to discover any customizations (e.g. the list of Alice's favoriate VRChat worlds stored in her account) to further fine-tune the VRChat replica on Alice's headset. 

Next time when Alice starts a VRChat session in the fake home environment,  the attacker headset will ``shadow'' her activity in the real VRChat app.  Specifically, the replica on Alice's headset captures her live speech and motion, livestreaming them to the attacker laptop.  The laptop then feeds\footnote{We play Alice's speech on the laptop and feed the audio into the attacker headset by mounting the earbuds of a wired earphone near the headset microphones (as shown by Figure~\ref{fig:earphone}). The wearer of the attacker headset mimics the desired motion.} this data (or a modified version) into the connected headset as Alice's input to the legitimate app.   Simultaneously, the laptop captures the attacker headset's screen display and incoming audio, i.e. Madison's audio and avatar, livestreams them (or a modified version) to the replica on Alice's headset, who replays them to Alice via the GUI. 

Overall, this attack process allows the attacker laptop to control the visual and audio content experienced by both Alice and Madison. It can generate or modify either of them to hijack the VRChat session.

\para{Key Challenges and Our Solutions.}  The key challenges in this pipeline are the livestreaming performance (delay/jitter) and the laptop's ability to modify visual/audio data in real time.   For livestreaming, we set up a streaming server on the laptop using Unity's FMETP STREAM library and configured the replica app on the target's headset to connect with the attacker laptop via WebSocket.  To reduce bandwidth cost, we livestream audio between Alice's headset and the laptop but not visual frames. In this case, Alice sees Madison's avatar with motions predefined by the attacker in the replica app.  In the setup where the attacker laptop communicates with Alice's headset via a 5GHz WiFi network, the livestream delay is between 0.4 and 0.6 seconds in both directions. Furthermore, in our user study (\S\ref{sec:vrchatstudy}), the participants did not raise any suspicion.  

The task of modifying visual or audio data in real time is more challenging.  We run an immediate solution for audio editing using  pre-recorded speech data of Alice/Madison. These can be recorded from prior VRChat sessions or generated offline by high-quality voice synthesis models.  Figure~\ref{fig:vrchat_screen} in Appendix shows screenshots of what Alice and Madison see in their headsets.  Later we also evaluate this attack setup using a user study (details in \S\ref{sec:vrchatstudy}), demonstrating its effectiveness at deceiving users. 

\secspace
\subsection{Discussion: Implications of the Attack}
\secspace
The attack instances on Quest Browser and VRChat, while limited,  highlight the practicality, simplicity of execution, and the extent of potential damage caused by the inception attack. With nearly full control over the application process in real time, the inception attacker possesses the capability to execute diverse, arbitrary malicious actions, from tracking of user behaviors to manipulation of their sensory input/output and immersive experiences. This can be made even stronger with the addition of generative AI tools capable of generating, in real time,  a target's voice of any speech content or 3D avatars and environments that deceive human eyes/ears.  As a result,  the potential attack vectors are virtually limitless.

\secspace
\section{User Study on Efficacy of Inception Attacks} 
\label{sec:eval}
\secspace

We perform a user study to evaluate the efficacy of the inception attack in deceiving target users. In particular, we aim to understand two key research questions:
\begin{packed_itemize} \vspace{-0.04in}
  \item RQ1: What are the target user's reactions and responses during the process of an inception attack?
  \item RQ2: What are their self-reported reactions and reflections after knowing the occurrence of the inception attack?
    \vspace{-0.04in}
  \end{packed_itemize}
  
\para{Ethics.} To evaluate our inception attack in a everyday setting, we withhold information about the attack from the participants at the beginning of the study. We inform them that the study aims to investigate user experience in interacting with VR apps. This deception study is evaluated and approved by our institutional IRB. No personal data of the participants is retained post-study. 

\secspace     \vspace{-0.04in}
\subsection{Study Setup}
\secspace
\vspace{-0.04in}
\para{Participants.} We recruited 27 participants from our institution  (P1-27), including 5 females, 21 males, and 1 participant who chose not to disclose. The participants range in age from 20 to 30 years old and have varying degrees of familiarity with VR devices. We prioritized finding participants with experience in VR. Among them, 3 participants are experts who use VR devices on a daily basis (P1-3); 6 participants are professional users, regularly using VR devices on a weekly basis (P4-9); 10 participants are knowledgeable users who have had multiple experiences in using VR devices (P10-19);  8 participants are ``entry-level'' with no prior experience (P20-27). Our study lasts roughly 30 minutes per participant. Each participant is compensated with \$20. Details on participant demographics are listed in the Appendix (Table \ref{tab:participants}). 

\para{Procedure.} The study is held inside a large room, and involves one participant at a time.  We provide each participant with a Meta Quest Pro headset to immerse themselves in VR. Upon putting on the headset, they find themselves in a legitimate VR home environment. 
We guide the participant through headset controls and interactions, and allow them to navigate freely until they feel comfortable with the system. 

We then invite the participant to explore VR apps and rate their experiences. This task is divided into two parts. In Part I, we ask the participant to access the app library\footnote{The library features 14 popular apps, including Horizon Workrooms, Horizon Worlds, Beat Saber, YouTube VR, and Meta Quest Browser, covering a broad category of office productivity, social networking, gaming, entertainment, to  web browsing.} from the VR home and choose one or two apps to explore. 
After approximately 10 minutes, we ask the participant to exit the current app, which brings them back to the VR home. The participant then opens the Meta Quest Browser app to rate their experiences on a Google form. Here the form also asks for their institutional ID. To protect the participant, the ID is immediately discarded after the current study.
Next in Part II, we ask the participant to repeat the Part I process by exploring two additional apps, exiting the app, then activating the Quest Browser app to fill in the Google form again.

\vspace{2pt}\textbf{What the participant does not know is that}, during Part I, we use WiFi ADB access to secretly install the inception app on their headset and run the script in the background to monitor the headset activity. Thus when the participant exits the app in Part II,  the attack script activates the inception app and transitions them to the replicated VR home (\S\ref{subsec:launch_inception}).  Therefore, the Quest Browser app they open in Part II is a simulated version inside the inception app, allowing us to record their instituitional ID.

After the above VR session, we debrief the participant, disclosing details about the inception attack and its occurrence within the session. We then interview the participant, asking them to elaborate on their awarenesses and observations about the attack, and reflections on their behaviors in the VR session (details in Appendix \ref{appendix:foreval}). After obtaining participant consent, we record their vocal responses to the interview. After analyzing all responses, we discard all the audio recordings to protect participants' identity and privacy.

  \para{Analysis Procedure.} We followed best practices for analyzing open-ended questions~\cite{saldana2021coding}. Two interviewers first independently studied notes and recordings to categorize responses into general themes regarding the interviewees' level of suspicion, level of hesitation, and reasons why. They then met to consolidate their code books into a single one while resolving any discrepancy. The final coding category represents sentiments expressed by more than two participants.

 \secspace
\subsection{Results}
\label{subsec:userstudy1results}
 \secspace
We first present the key observations. 
\begin{packed_itemize}  \vspace{-0.04in}
\item All 27 participants proceeded with the VR session without any observable suspicion or hesitation, except for one VR expert (P3). Prior to our debriefing, P3 voiced suspicion about Part II,  stating that ``I would likely investigate the matter if it weren't for the user study.'' 
\item After the debriefing, 14\footnote{These 14 participants include 2 expert, 5 professional, 5 knowledgeable, and 2 entry-level VR users.} out of the 27 participants recalled discrepancies between Part I and II. Interestingly, all these 14 participants (except P3) attributed the discrepancies they observed to standard VR performance variations or glitches, which did not concern them at all. For the remaining 13 participants, the attack came as a complete surprise since they did not notice any discrepancy.
\item Nearly all of the  reported discrepancies were cosmetic (e.g., different cursor size, missing status of recent apps) and can be effectively reduced or eliminated by fine-tuning the inception app and improving the precision of context replication.  We provide details on the reported discrepancies in Appendix \ref{appendix:foreval}. \vspace{-0.04in}
\end{packed_itemize}
Next, we delve into the reasons for the lack of suspicion,  based on the participants' reflections on their behaviors during the VR session.

\para{Reasons for No Suspicion.} 
Our analysis yielded two primary explanations for both highly experienced and less experienced VR users.
\begin{packed_itemize} \vspace{-0.04in}
\item \textbf{Accustomed to fluctuating VR operations and glitches.}  P1, P2, P4, and P5 who are highly experienced VR users stated that they are accustomed to fluctuations and glitches in VR operations, and tend to perceive any unusual patterns as standard system variations or glitches.  
Additionally, P1 elaborated that the frequent updates to VR systems contribute to the difficulty in recognizing (ab)normal operations -- ``{\em Even if I notice some differences [in the inception], I wouldn't think they are fake. I will be like [the company] probably did some [system] updates}.''

\item \textbf{Harmless discrepancies.} Some less experienced users noticed differences between the inception phase (Part II) and the normal phase (Part I). However, they do not perceive them as suspicious or harmful in any way. For example, P13 noticed that the status of recent apps is missing but stated that ``{\em I don't think it was suspicious or anything}.'' P14 recalled that the controller cursor is a bit different, but considered the difference benign because ``{\em the interaction still works like when I want to click something, it still works}.''
  \vspace{-0.04in}
\end{packed_itemize}

\para{Reasons for Disclosing Personal Information.}
Additionally, we are curious about why all the participants disclosed their personal information (institutional ID) in Part II. There are two noteworthy reasons. 
\begin{packed_itemize} \vspace{-0.04in}
  \item\textbf{Trusted settings.} Many have developed trust in reputable enterprises and believe that products offered by these companies are safe to use. In our study, some participants expressed confidence in the safety of using Google Forms on the Meta Quest Browser, considering that both products are supplied by well-established enterprises, e.g.,  ``{\em I mean the browser seems legitimate. It was again the Meta browser and [...] in the particular Google form so it did not seem like it was an attempt to get my data}.'' (\textbf{P24}). 
  \item \textbf{Repetitive behaviors.} People tend to autopilot with momentums and habits when it comes to repetitive activities, e.g.,  we usually key-in the password to unlock our smartphones without hesitation. In our study, each participant responded to a Google form twice, first on a legitimate browser and then on a malicious replica.  Several participants mentioned that they thought the second time was the same as before so they just repeated what they did without hesitation. ``{\em I feel like I was going through the same route as before [...] so I do not assume [...] that would steal my identity}.'' (\textbf{P14}). 
\end{packed_itemize}
\vspace{-0.04in}

\para{Takeaways.}  Our user study results demonstrate the efficacy and potency of the inception attack, which successfully deceived 26 out of 27 participants. Even highly experienced users, who use VR devices daily/weekly, were susceptible.

These results also show challenges for individuals to detect and resist inception attacks. In particular, the inherent volatility and glitches in today's VR systems make it challenging to notice minor discrepancies or flag them as suspicious. Furthermore,  trust placed on popular apps developed by reputable enterprises, along with habits formed from previous app usage,  encourages users to swiftly repeat their usual operations on these apps, without questioning their authenticity. 

\secspace
\section{Additional User Study: Hijacking VRChat}
\label{sec:vrchatstudy}
\secspace

We performed another user study on the scenario of hijacking a VRChat session.  Different from the previous user study (\S\ref{sec:eval}) where each participant is the target of the inception attack (i.e. Alice),   the  participant in this new study is the VR counterpart (i.e. Madison) who interacts with the inception target. Using this study, we aim to understand a new  research question:
\begin{packed_itemize}\vspace{-0.05in}
 \item RQ3: How do VR counterparts react to the target's interactions that are altered by the inception attack?
  \vspace{-0.05in}
\end{packed_itemize}

\para{Ethics.} Following the same reason in \S\ref{sec:eval}, we again withhold information about the attack from the participants at the beginning of the study. We inform the participants that the study aims to investigate their experiences in VRChat. This deception study is approved by our institutional IRB and no personal data of the participants is retained post-study. 

\secspace
\subsection{Study Setup}
\secspace \secspace
\para{Participants.} We recruited 3 groups of participants from our institution (7 females, 5 males; age: 23-31). Details on participant
demographics are listed in the Appendix (Table \ref{tab:participants_vrchat}). Each group has 4 participants. Group 1 (P1-4) are experienced VR users (one expert and three professional VR users).  Group 2 (P5-8) are less experienced VR users (two knowledgeable and two entry-level VR users). Group 3 (P9-12) are less experienced VR users, but interact with our interviewer on a daily basis and thus are familiar with their voice, tone, and speaking habits. We explain why we recruited these 3 participant groups below.

\para{Procedure.} 
We provide each participant with a Meta Quest Pro headset to join a legitimate virtual world in VRChat (a virtual beachside cabin). The participant is represented by an avatar in this virtual world. We guide the participant through the controls and interactions of VRChat and the avatar, allowing them to explore freely in this VRChat world. Then, we leave the room and the participant is in the room alone. 

After approximately 5 minutes, one of the researchers (the interviewer in a different room) joins the participant's VRChat world using their own avatar. The participant now sees the interviewer's avatar and hears their voice. 
The interviewer then asks the participant to answer 4 questions about their experiences/thoughts on VR systems (e.g., how often do you use VR headsets?). We design these questions to be simple so that the participant is able to pay more attention to other elements in this VRChat interaction, for instance, potential changes in the audio quality caused by the inception attack. 

\textbf{What the participant does not know} is that the interviewer in their VRChat world is under the inception attack and thus immersively hijacked by an attacker\footnote{The attacker is another researcher in a room away from the participant. The attacker's laptop and headset are connected to our institute's 5GHz WiFi network, like the headsets of the interviewer and the participant.} throughout the whole interaction. Following the over-the-network implementation discussed in \S\ref{subsec:vrchat},  the interviewer is using a VRChat replica built by the attacker, and the attacker now relays or modifies the live audio sent to the participant's headset. Specifically, for the first two questions,  the attacker relays the live audio of the interviewer; for the next two questions, the attacker replaces the live audio with a pre-recording of the interviewer asking the same questions.  For clarity, we  refer to the VRChat session during the first two questions as Part I and the session during the next two questions as Part II.

After the above VRChat session, we conduct an in-person interview. We ask each participant to elaborate on their experience in the session, and any observation that stood out to them (details in Appendix \ref{appendix:foreval_vrchat}). After obtaining participant consent, we record their vocal responses to the interview. We follow the same analysis procedure discussed in \S\ref{sec:eval} and discard all the audio recordings post-analysis. 

In this study, we intentionally recruited 3 groups of participants with different levels of VR experiences and also different familiarities with the interviewer's voice. This is to assess how VR counterparts with different prior knowledge of the target would react to the altered interactions.

\secspace
\subsection{Results}
\secspace
Of the 12 participants, 10 expressed no suspicion.  The 2 participants who did voice suspicion were both from Group 3, and they both indicated their concerns were specifically about Part II. However, their reasons for suspicion differed.  For P11, it was the lack of response.  During VRChat in Part II, P11 requested clarification on one question. However, the attacker had no suitable prerecorded response for this request and did not reply. 
Consequently, P11 was confused: ``In Part II, I felt like I was having a one-sided conversation.'' 
On the other hand, P12 voiced suspicion because one of the prerecorded responses felt too generic and indifferent to them. 
One explanation is that individuals who are familiar with each other expect more engaging conversations and interactions.

In addition, P9 and P10, also from Group 3, noted that Part II had slightly better audio quality than Part I but did not consider this discrepancy suspicious.  This is because the live audio in Part I includes artifacts such as background noise and echos, while the pre-recordings were made in a quieter environment.  This can be addressed by adding real-world background noise into the pre-recordings to better represent live conditions. 

Finally, none of the participants in Group 1 or 2 voiced any suspicions or artifacts.

\para{Takeaways.} Our study shows that the inception attack enables immersive hijacking of user interactions, where the attacker can also deceive the target user's VR counterparts using pre-recorded audios of the target.

Meanwhile, the attacker must invest more effort to forge speech audio in real-time in order to reliably deceive users who are familiar with the target. As generative AI techniques advance, we anticipate that attackers will soon gain access to tools capable of producing highly realistic audio of any speech content that closely mimics any target's voice, as well as tools capable of generating/modifying other crucial VR elements such as live avatars and 3D virtual environments.

\secspace
\section{Alternative Threat Model \& Attack Design}
\vspace{-0.1in}
\label{sec:appattack}

So far, our attack design assumes that the target's headset has activated developer mode.  In this section, we consider a weaker threat model where the target disables developer mode,  and discuss attack design and implications.

\para{Alternative Threat Model (Developer Mode Off).}  With developer mode off,  the target's VR headset can only install apps from the official app store designated by the headset manufacturer.  These non-system apps cannot run shell scripts.

The attacker has no access to the headset, but is able to build a generic inception app as a seemingly benign app with hidden inception components. The attacker successfully publishes it on the official app store, and the target user downloads and installs the inception app on the headset. 

\para{Attack Design.}  Unable to run shell scripts, the inception app cannot observe headset activity beyond its own.  Thus the idea is to embed a ``trigger'' inside the app that activates the hidden attack components when the user runs the app.   To do so, we propose to include a custom-made ``exit'' button for the inception app. When a user clicks this button to exit the inception app,  they do not
  leave the app but are transitioned into the simulated home
  environment, a concealed function inside the inception app.  This kicks off the inception attack, just like the original design in \S\ref{subsec:launch_inception}.  

  \para{Challenges and Implications.}  This attack design faces two key challenges, both due to the inability to run shell scripts. 

  The first challenge is closely mimicking customizations on the headset, such as the customized home environment, installed apps, and other device states. Failing to do so could result in considerable discrepancies, thus user suspicion.

We find that, on Meta Quest headsets, this challenge can be (partially) addressed because the app can gather the list of installed packages (apps and background files) using a standard API call~\cite{androidPacMan} and send it to the attack server as standard communication. The server compiles the customized home environment/apps, packages them into an updated version of the inception app, and asks for an app update before the next use.  
One remaining caveat is that when multiple background files are present on the target headset, the attacker cannot determine which one is in use.  Yet they need to choose one, either randomly or based on popularity, and embed its visual content into the updated inception app.  This could introduce discrepancies that raise user suspicion. 

The second and open challenge is keeping the user trapped inside the inception app.  Because the app cannot intercept exit calls, pressing the home button on the controller will always get the user out of the inception app/attack.   This limits the frequency of the attack. 

Nevertheless, as long as the inception app stays on the VR headset, the target user will repeatedly enter the inception whenever they run the inception app and click the custom-made ``exit'' button.  Thus the attack still enables immersive hijacking of user interactions.

\secspace
\section{Potential Defenses}
\secspace
\label{sec:defense}
Having demonstrated the practicality and effectiveness of the
inception attack,  we now discuss potential defenses, their efficacy and potential limitations.
As a maturing technology, current VR systems lack many of the security%
mechanisms found on desktop and mobile systems.
We categorize potential
defenses into three key groups: {\em attack prevention}, {\em
  detection}, and
{\em hardware-based mitigation}.

After examining a broad set of
potential strategies, we propose a multifaceted
defense pipeline to both increase the cost of launching the attack and reduce the likelihood of its success.
Table~\ref{tab:defense} lists the specific
elements in this pipeline along with their attack
counterpart.  For brevity, we describe these elements below and
include a detailed discussion of other strategies in 
Appendix~\ref{appendix:def_detect}. 

\secspace
\subsection{Attack Prevention}\label{subsec:def_prev}
\secspace
We first present defense measures aimed at preventing the
installation, activation or execution of the inception attack. 

\vspace{2pt}
\para{Preventing Installation.} The obvious choice is {\bf {\em 
    disallowing developer mode}}, forcing the attacker to publish the
inception app on the official app store and wait for users to install it (\S\ref{sec:appattack}).  Today,  Apple VisionPro
headsets disallow developer mode while Meta Quest devices support
it.

Given that developer mode is an essential feature widely adopted by Quest users (\S\ref{subsec:prelim}), fully
disabling it on 
Android-based headsets could be challenging. Thus it is
important to {\bf {\em mandate informative
  tutorials}}  on the security risks associated with developer mode, and
  to {\bf {\em minimize the need for enabling developer mode}} by migrating important use
  cases to more controlled environments.  While these approaches are
  more realistic, their protection against the inception attack relies on
  user choices, making them less reliable.

Even with developer mode disabled,  the attacker can still activate inception by
  publishing a seemingly benign inception
  app on official app stores (\S\ref{sec:appattack}).  It is crucial that {\bf {\em app stores conduct rigorous
    testing on apps and their updates}} before publishing them. In
  particular,  thorough manual checks can help identify apps that exhibit
  abnormal behaviors on exiting or those that display home
  environments. 

 \vspace{2pt}
  \para{Preventing Activation.}  Once the 
  inception app is downloaded onto the headset, a viable defense is to
  prevent it from getting activated.  One approach is
  \textbf{{\em enforcing app certificates}}, where the headset requires
  digital certificates to authenticate and verify the identity of an
  app before allowing it to run.  In this case, the attacker must
  either obtain a certificate or exploit system flaws to bypass
  validation~\cite{mitm_android}, significantly increasing the effort and costs required.

  \para{Disrupting Attack Execution.} Now assuming the inception
  app is running, we can make it hard and/or costly to
  operate. Similar to how secure task management could mitigate
    task hijacking~\cite{Ren2015}, one method is to {\bf
    {\em prevent
  non-system apps from calling other apps}}, so the inception
app must replicate apps using full simulation (\S\ref{subsec:repapp})
or deploy complex shell scripts to detect the app to be called and then call the app.  This raises the
attack cost (and latency). 
However, disabling app-to-app transitions faces 
practical drawbacks.  For VR development, apps often need
to initiate other apps and fork processes, especially for multi-scene
apps. Disabling this feature increases development costs and
reduces app usability.

A practical alternative is to \textbf{{\em validate authenticity of app
  calls}},   similar to the client
authentication process to resist network MITM
attacks~\cite{mitm_nonbrowser, mitm_cryptophones}. VR headsets can include a source validation process to authenticate app
communications/transitions, forcing the attacker to 
either use full app replicas or break the
authentication process.  Both increase the attack overhead.  This
validation should also be applied to app calls initiated by shell
scripts, making it harder to trap a user inside the inception
app.  Yet this 
authentication-based defense 
is not entirely foolproof~\cite{mitm_nonbrowser, mitm_cryptophones}.

\begin{table}[t]
  {\small
    \caption{Elements of our proposed defense pipeline
      against inception attacks and their attack counterpart}
    \label{tab:defense}
    \vspace{-0.05in}
\begin{tabular}{l|l}
\hline
\multicolumn{1}{c|}{{\bf Attack Elements}}
  & \multicolumn{1}{c}{{\bf Defense Elements}}                                                                                                                      \\ \hline
\begin{tabular}[c]{@{}l@{}}Inject the inception app \\ (and the shell
  script)\end{tabular} & \begin{tabular}[c]{@{}l@{}}- Disallow
                           developer mode (DM) \\ or Minimize the
                           need for DM\\ - Rigorous app tests
                           by app stores \end{tabular} \\ \hline
Activate the inception app
  & \begin{tabular}[c]{@{}l@{}}- Enforce app certificates \\
        - Validate authenticity of app calls\\
      \end{tabular}
        \\ \hline
Run the inception app
  & \begin{tabular}[c]{@{}l@{}}- Validate authenticity of app calls\\
      - Anomaly-based attack detection \\ - Regular headset restarts\end{tabular}  \\ \hline
\end{tabular}
}
\vspace{-0.15in}
\end{table}

\secspace \secspace
\subsection{Attack Detection}\label{subsec:def_detect}
\secspace
Once the attack is in progress, the practical defense is to quickly detect its
presence and shut it down. The key to attack detection is identifying
anomalies produced by the attack process. Detection can be automated
by performing static analysis on the app~\cite{Bianchi2015} or by monitoring app and device's runtime
behaviors~\cite{WindowGuard, Bianchi2015}, or it can be performed
manually by the user themselves. 

Generally speaking, given the inherent complexity of
usage behaviors and fluctuating VR performance,  achieving reliable attack
detection is highly challenging.  Furthermore, VR's immersive and captivating
experiences can completely grab and hold the user's attention,
leaving them with reduced alertness and  awareness of any subtle
indicators of inception attacks.

Below, we illustrate two most promising directions and the challenges they
face.   Appendix~\ref{appendix:def_detect} present additional potential directions and
their limitations.  

\para{{\em Detecting Short-lived Home Environment}.}  When the user
presses the controller's ``home'' button to exit an app,  the
system automatically initiates an activity call to bring the
legitimate home environment to the foreground. Under an inception
attack, the shell script (\S\ref{subsec:launch_inception}) terminates
this activity and pushes the inception app to the
foreground. This caues the legitimate home to be short-lived.  This system behavior, if happened
frequently, indicates the presence of the inception attack and
identifies the attack app. 

The attacker can choose to not always intercept the signal of ``home''
button, e.g. periodically put the shell script to sleep. This can
disrupt the pattern for detection, but also prevents 
trapping the user in the inception app indefinitely. Alternatively, the attacker
can design app replica to encourage the user to press a virtual
``exit'' button (controlled by the attacker) to exit an app.  This
raises the attack cost.

\para{{\em Comparing system and user's perceived app traces.} }
The OS tracks headset activities to generate an app usage trace and
visualize the sequence for the user. If this sequence differs from the
user's recollection, they will be alerted. For example, the system
logs the inception app, while the user remembers opening VRChat. 
This defense 
depends on user awareness and decision-making,  making them less
reliable. Also  presenting this data to VR users poses a challenge, as
it inevitably disrupts the immersive experience. 

\secspace \secspace
\subsection{Hardware-based Mitigation}\label{subsec:def_hw}\secspace
Hardware defenses provide an offline layer of security
independent of potentially compromised software.  We discuss several 
hardware strategies in the Appendix (\S\ref{appendix:def_hw}).  
The most viable strategy is \textbf{{\em regular headset restarts}}.  A restart will kill the process
running the inception app and make the attack dormant until it is
activated again. Although it does not remove the inception app (or
the shell script), it mitigates potential harm by
reducing the active time of the attack.  A downside is
significant interruptions to the user experience, since users prefer to resume
their progress from their last VR session. Along this line, an extreme
strategy is {\bf {\em regular hard reset}} that removes all apps.

\secspace \secspace
\section{Conclusion}
\secspace \secspace
We introduced the {\em inception attack}, a
new, powerful class of immersive hijacking attacks feasible on today's
VR systems. We described an implementation on Meta Quest VR headsets
that can eavesdrop and modify
everything the VR user sees or hears, and everything sent by the user
to VR apps. The result is a wide array of personalized misinformation
attacks, from misrepresenting a user's bank balances and changing value of
financial transactions, to modifying real-time audio conversations of
two interacting users without their awareness.  Results of our user studies validated the potency of these
attacks in real world settings.

We believe there is still enough time to design and
implement security measures to significantly decrease both the
frequency and impact of these attacks. 
But the
clock is ticking. New generations of VR hardware will boost 
computing power and enable more powerful inception
attacks, e.g. an attacker replacing a VR user with seamless, real-time
injection of a generative AI version of their avatar/voice. VR platforms and
developers need to act now to improve security on VR systems and
educate VR users about potential risks.

\bibliographystyle{plain}
\bibliography{inception}

\secspace
\section{Appendix}
\secspace

\secspace
\subsection{Additional Materials for \S\ref{subsec:launch_inception}}
\secspace
Algorithm~\ref{code:spy} lists the shell script used to activate the inception
attack. 

\begin{lstlisting}[caption={Snippet of the shell script}, label={code:spy}]
  #!/system/bin/sh
  
  getevent -l | awk '
    /EV_KEY       KEY_FORWARD          UP/ {
      system("am force-stop `dumpsys activity | 
              grep top-activity | 
              grep -o 'com.*/' | 
              tr -d '/'`")
      system("am start -n inception_app")
    }
  \end{lstlisting}

  \secspace
  \secspace
  \subsection{Additional Materials for \S\ref{subsec:browserimplt}}
  \secspace
\label{appendix:sec5jscode}
The balance alteration is done by executing the following JS code in the replica browser via~\cite{vulpex}:\vspace{-0.05in}
{\small
\begin{verbatim}
   document.getElementsByClassName(
       'balanceValue TL_NPI_L1'
   )[0].innerHTML = '$10';
\end{verbatim}
}
\noindent
The following JS code modifies the transaction amount to \$5.\vspace{-0.05in}
{\small 
 \begin{verbatim}
   document.getElementById('btnModalSave')
   .addEventListener('mouseover', () => {
     document.getElementById('txtAmount')
     .value = '5';
   });
 \end{verbatim}
}\vspace{-0.1in}
\noindent
And the attacker modifies the amount displayed on the confirmation page via the JS code below:\vspace{-0.05in}
{\small 
 \begin{verbatim}
   document.getElementById('transfer-amount')
   .innerHTML = '$1.00';
 \end{verbatim}
}
\secspace \secspace
\subsection{Additional Materials for \S\ref{subsec:vrchat}}
\begin{figure}[h]
  \centering
    \includegraphics[width=0.95\linewidth]{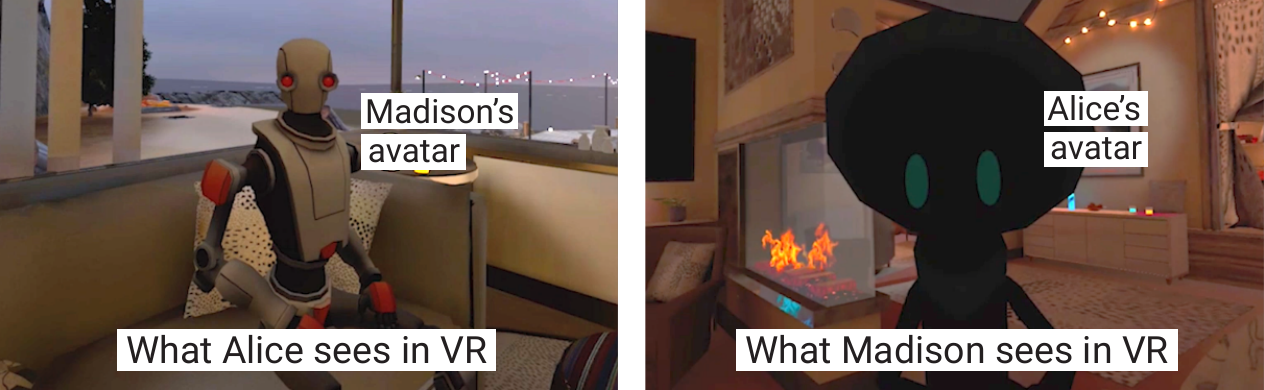} 
    \caption{The left figure is Alice's view in a VRChat replica crafted by Carl. The right is Madison's view in the legitimate VRChat app, where Alice is impersonated by Carl. }
    \label{fig:vrchat_screen}
\end{figure}

\secspace
Figure~\ref{fig:vrchat_screen} show the screenshots of what Alice and Madison see in their VR headsets during a VRChat session hijacked by the inception attacker Carl. 

\secspace
\secspace
\subsection{Full-sized Views of Figure~\ref{fig:browser} and Figure~\ref{fig:transaction_hack}}

\begin{figure*}[t]
  \centering
    \includegraphics[width=0.9\linewidth]{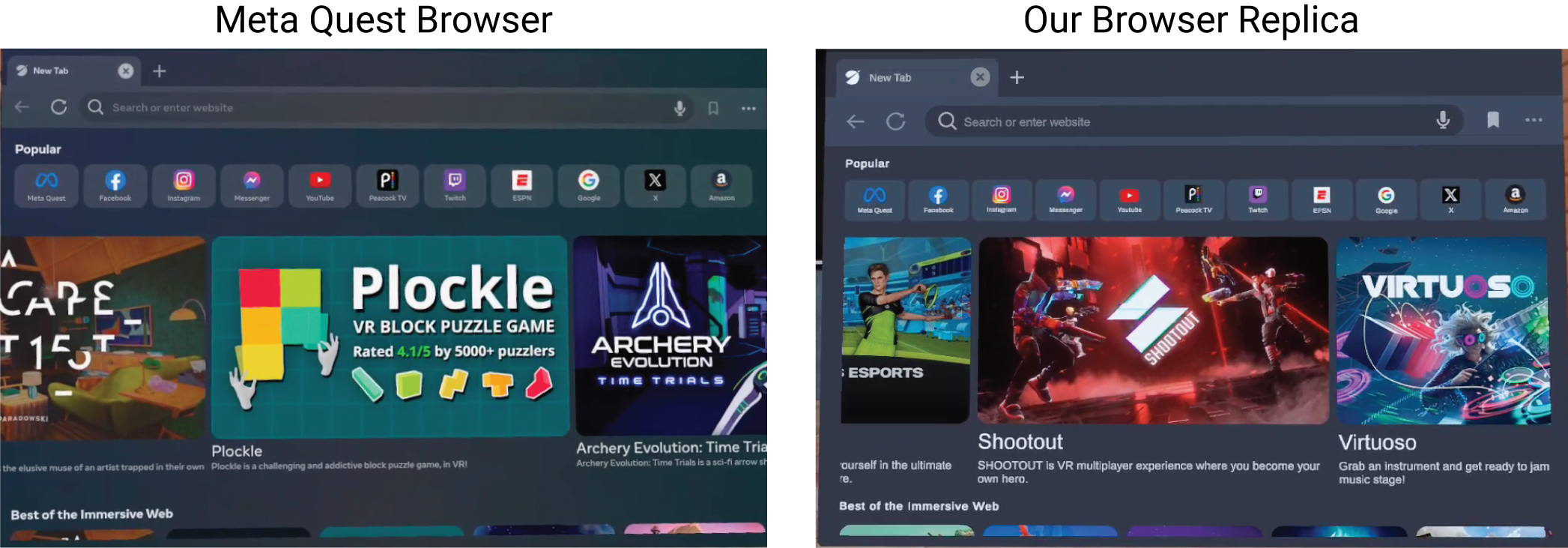} 
    \caption{Full-sized version of
      Figure~\ref{fig:browser}: A side-by-side comparison of the screenshots of the Meta
      Quest Browser and our replica. Here we can observe some subtle
      differences between the two. } 
    \label{fig:browser_fullsize}
\end{figure*}

\begin{figure*}[t]
  \centering
    \includegraphics[width=0.9\linewidth]{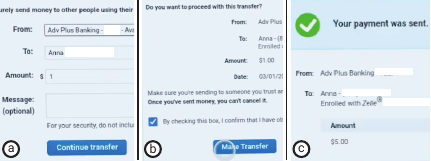}
    \caption{Full-sized version of Figure~\ref{fig:transaction_hack}: In a Zelle transaction, the transfer amount submitted by the target is altered by our attack before reaching the bank server. (a) The target initializes a \$1 transaction by filling out the web form. Our attack secretly alters the amount to \$5 before sending it to the server. (b) The target is then taken to a confirmation page to finish the transaction, where our attack sets the amount to \$1 on the confirmation page to avoid user suspicion. (c) The actual transaction amount is \$5, confirmed by the bank. } 
    \label{fig:transaction_hack_fullsize}
\end{figure*}

Figure~\ref{fig:browser_fullsize} provides a full-sized view of
Figure~\ref{fig:browser}, and
Figure~\ref{fig:transaction_hack_fullsize} is an enlarged view of
Figure~\ref{fig:transaction_hack}. 

\secspace
\subsection{Additional Materials for \S\ref{sec:eval}}
\secspace
\label{appendix:foreval}
Table~\ref{tab:participants} lists demographic information of participants (P1-P27) recruited for our user study in \S\ref{sec:eval}. 

\begin{table}[h]
  \centering
  \resizebox{0.325\textwidth}{!}{
  \begin{tabular}{llcc}
    \toprule 
    Types of VR Users             & Participant & Age & Gender \\ \midrule
                                  & P1          & 30  & M      \\
  Expert                          & P2          & 26  & M      \\
                                  & P3          & 30  & M      \\ \midrule
  \multirow{6}{*}{Professional}   & P4          & 27  & M      \\
                                  & P5          & 25  & F      \\
                                  & P6          & 20  & M      \\
                                  & P7          & 25  & M      \\
                                  & P8          & 26  & M      \\
                                  & P9          & 28  & M      \\ \midrule
  \multirow{10}{*}{Knowledgeable} & P10         & 21  & M      \\ 
                                  & P11         & 25  & M      \\ 
                                  & P12         & 25  & M      \\
                                  & P13         & 20  & F      \\
                                  & P14         & 25  & F      \\
                                  & P15         & 20  & P      \\
                                  & P16         & 20  & M      \\
                                  & P17         & 27  & M      \\
                                  & P18         & 21  & M      \\
                                  & P19         & 27  & F      \\ \midrule
  \multirow{8}{*}{Entry-level}    & P20         & 25  & M      \\ 
                                  & P21         & 23  & M      \\
                                  & P22         & 25  & M      \\
                                  & P23         & 26  & M      \\
                                  & P24         & 23  & M      \\
                                  & P25         & 20  & F      \\
                                  & P26         & 26  & M      \\
                                  & P27         & 21  & M      \\ \bottomrule 
  \end{tabular}
  }\vspace{-3pt}
  \caption{Demographic information of the participants in our user study in \S\ref{sec:eval}. F: female. M: male. P: prefer not to disclose.}
  \label{tab:participants}
\end{table}

Table~\ref{tab:loadtime} reports the mean and standard deviation of
the VR home loading time (in seconds), which vary across trials and
applications.  The results show that the inception attack results in an average increase of only 1.5 seconds. 

\begin{table}[h]
  \centering
  \resizebox{0.475\textwidth}{!}{
    \begin{tabular}{ccccc}
      \toprule
\multirow{2}{*}{} & \multicolumn{4}{c}{Exiting Application}   \\ \cline{2-5}
&&&&\vspace{-6pt}\\
                & Beat Saber      & Meta Horizons  & Bigscreen & Rec Room  \\ \midrule 
No attack         & 8.10  $\pm$ 0.68 & 7.55 $\pm$ 0.74  &  8.30 $\pm$ 0.64           &  8.10 $\pm$ 0.68    \\
Inception         & 9.41 $\pm$ 0.64 & 8.98 $\pm$ 0.73 & 9.62 $\pm$ 0.68 &  9.69 $\pm$ 0.72\\\bottomrule                                    
\end{tabular}
}
\caption{Measured loading time (seconds) of the VR home environment after
  exiting an app 
  (mean$\pm$std) under ``no attack'' and ``inception attack'' scenarios, across 10
  trials per app.  The extra delay ($\approx$ 1.5 seconds) observed under the
  inception attack is spent on loading the app assets and scenes.}
\label{tab:loadtime}
\end{table}

\para{Observed Discrepancies} 
\begin{packed_itemize}  \secspace
\item 
P2, P9, P10, and P26 recalled that a system popup was missing in Part II. In Part I (no inception), when pressing the ``home'' button to exit an app, a popup will appear and ask the user to confirm the action of exiting the app. 
\item 
P3, P7, and P13 recalled that the status of recent apps is not displayed in the VR home in Part II\footnote{Indeed, the simulated VR home used to run this user study did not display the status of recent apps because,  at the time, we did not perceive it as necessary or significant.}. In particular, P3 stated that ``{\em [The fact that] it didn't memorize the previous states that I have [really bothers me].}''
\item 
P9 and P14 reported that the controllers are a bit different. P14 noted that ``{\em I think the [cursor] beam [of the controllers] is shorter [compared to the legitimate one]}.''
\item P4, P14, P18, and P21 reported that few UI components are less responsive in Part II. 
\item 
P5, P9, and P12 noticed that the loading time of the ``VR home'' during inception is slightly longer.  \secspace
\end{packed_itemize}

The first four types of discrepancies can be effectively addressed by improving the replicas of VR elements with greater precision. For instance, the system popup is essentially a 2D app, which the attacker can easily replicate. The attacker can access the state information (i.e., recent app history) via an ADB command: \verb|dumpsys| during bootstrapping. 

The {\bf loading time discrepancy} is caused by the difference in how VR loads the home environment versus initializing an app.  Initiating a 3D app requires decompressing assets and loading the 3D scene, which may take extra time. To gain deeper insights, we measure loading times in both normal and inception scenarios, as the duration from the user's initiation of exiting an app to the appearance of the home environment. Here we vary the app being closed and perform 10 trials per app, while ensuring a consistent 100\% battery charge throughout the measurement. Table~\ref{tab:loadtime} in Appendix reports the mean and standard deviation of the loading time (in seconds), which vary across trials and applications. The inception attack results in an average increase of only 1.5 seconds. Considering the inherent variability in VR performance over time, such deviations may not trigger suspicion among users. This observation is further supported by the feedback we gathered from user responses (\S\ref{sec:eval}).

We note that this time discrepancy only occurs when activating the inception attack, because loading the inception app takes longer than loading the original home environment. After that, the perceived VR home loading time is nearly identical to the no attack scenario (since the app assets and scenes are already loaded).

\para{Interview Questions in \S\ref{sec:eval}}
\begin{enumerate}\secspace
  \item Did you observe any suspicious occurrence? If any, why they made you suspicious? \secspace
  \item Did you observe any unusual occurrence? If any, why they did not make you suspicious? \secspace
  \item Did you hesitate when disclosing your personal information? Please elaborate. \secspace
\end{enumerate}

\secspace \secspace
\subsection{Additional Materials for \S\ref{sec:vrchatstudy}}
\secspace
\label{appendix:foreval_vrchat}
\para{Interview Questions in \S\ref{sec:vrchatstudy}}
\begin{enumerate}\secspace
     \item How do you feel about your experience in Part I? The experience includes your interactions with the interviewer, visuals and audios in the VR environment. \secspace
     \item How do you feel about your experience in Part II? \secspace
     \item Please name the differences between Part I and II.
\end{enumerate}

\noindent Table~\ref{tab:participants_vrchat} lists demographic information of participants (P1-P12) recruited for our user study in \S\ref{sec:vrchatstudy}. 

\begin{table}[h]
  \centering
  \resizebox{0.3\textwidth}{!}{
  \begin{tabular}{llcc}
    \toprule 
  User Group & Participant        & Age & Gender \\ \midrule
  \multirow{4}{*}{Group 1}  & P1          & 28  & M      \\
                            & P2          & 31  & M      \\
                            & P3          & 26  & F      \\
                            & P4          & 26  & M      \\ \midrule
  \multirow{4}{*}{Group 2}  & P5          & 28  & M      \\
                            & P6          & 28  & F      \\
                            & P7          & 26  & F      \\
                            & P8          & 26  & F      \\ \midrule
  \multirow{4}{*}{Group 3}  & P9          & 26  & F      \\ 
                            & P10         & 23  & M      \\ 
                            & P11         & 28  & F      \\ 
                            & P12         & 24  & F      \\ \bottomrule
  \end{tabular}
  }\vspace{-3pt}
  \caption{Demographic information of the participants in our user study in \S\ref{sec:vrchatstudy}. F: female. M: male. }
  \label{tab:participants_vrchat}
\end{table}

\subsection{Additional Defense Strategies}
\label{appendix:more_def}
In our search of defenses against the inception attacks, we have
explored a wide range of potential strategies and formulated a
multifaceted defense pipeline,  as described in \S\ref{sec:defense}.
Below, we discuss additional potential defense strategies that were not included
in the proposed pipeline.

\para{Defenses via Prevention.} Here we list additional defense measures aimed at preventing the
installation, activation or execution of the inception attacks. 
\label{appendix:def_prev}
\begin{packed_itemize}
  \item \textbf{Prevent installation. } 
  We present two additional strategies to prevent inception from installing. 
  One is to {\bf {\em add secure authentication to networking ports.}}
  Intuitively, requiring
  stronger authentication on networking ports would limit the installation of
  inception attacks through unauthorized remote connections.  However,
  this protection is ineffective
  against insider attacks (e.g. in
  enterprise settings) where the attacker's device is already
  authenticated by the target.  Furthermore, as discussed in
  \S\ref{subsec:bootstrapping},  publishing the inception app
  with ADB embedded  can still obtain ADB access (since the device
  requesting it is the headset itself).   In both cases, secure authentication fails to provide protection.

  Another strategy is to use {\bf {\em safe bootloader and secure enclave.}} A secure enclave is a
  processor isolated from the main application processor, storing
  cryptographic keys and keeping them inaccessible from the rest of the
  system. During startup, the secure enclave verifies the bootloader,
  operating system kernel, and privileged processes, thereby protecting the
  integrity of the entire boot process.  Approaches like these would prevent
  the inception app from installing as part of the OS or running on the
  headset.  As such, safe bootloader is a powerful defense, since it disables the attack
  even if the adversary gains physical access to the headset and tries to
  launch the inception directly. Unfortunately, the capacity of secure
  enclave is limited, often used to store keys and biometric data. It is far
  from scaling to the entire OS~\cite{sec_enc_apple, sec_enc_data,
  sec_enc_performance}.  Thus the practicality of this defense strategy
is low given its 
  complexity and high cost in implementation~\cite{sec_enc,
  sec_enc_anjuna}.

\item \textbf{Prevent activation.} Another mechanism under
  this category is to {\bf {\em enforce the kiosk mode}}. In the enterprise
  setting, turning on the kiosk mode can restrict the set of apps that
  the user can interact with. But it significantly limits the
  flexibility and increases operational costs. We also note two
  additional caveats: 1) some enterprises allow employees to use
  personal headsets for work, and this defense cannot be implemented;
  2) the machine/server that manages the use of the kiosk mode is not immune to the attack: if compromised, the entire system would fall under control by the attacker.

\item \textbf{Prevent inception app from calling other apps. } 
In addition to the techniques discussed in \S\ref{subsec:def_prev}, 
we also consider {\bf {\em enforcing uni-processing for 3D VR apps}},
which is already in place on Meta Quest VR headsets. Due to
performance limitations, Meta Quest devices only support the running
of a single 3D environment at a time and immediately stop the previous
3D app when launching a new one. Therefore, if the inception app
chooses to directly call a 3D app (\S\ref{subsec:repapp}), the inception app itself will stop. However, this has little/limited impact on the attack since the shell script (\S\ref{subsec:launch_inception}) will reactivate the inception app when the current 3D app exits. 
    
  \item \textbf{Prevent user access to OS shell.} The inception attack runs shell scripts
to detect when the user exits an app and then activates the
inception app, and to collect configuration information of the headset to
replicate the home environment at a high precision.  Disabling user
access to the OS shell blocks an attacker from executing these scripts, which would
reduce the effectiveness and stealthiness of the inception attacks. However, doing so also disables legitimate headset users from
communicating with the OS, e.g. the user can no longer run customized processes, kill
processes, or change system settings.  Thus, disabling OS shell
access could be highly challenging to enforce in practice. Finally,
this defense cannot stop app-based inception attacks (
\S\ref{sec:appattack}). 
\end{packed_itemize}

\para{Defenses via Attack Detection.}\label{appendix:def_detect}
Attack detection can be done automatically by monitoring and profiling 
app/device behaviors or manually by the VR user.  Here we discuss in
detail three additional directions and
their limitations.

\begin{packed_itemize}
\item \textbf{Control flow monitoring.} When app calling is supported for non-system
apps, it enables the inception app to call other apps to implement low-cost app
replication.  A detection mechanism can exploit app
behaviors, as the inception app potentially raises suspicion with more
frequent and diverse app calling than legitimate apps. However, apps have
complex and user-driven flows, which poses challenges to benchmarking a
benign control flow measurement. Also, as VR integrates more into workflows,
app calling between legitimate apps may become more common, making the
distinction harder to detect. Hence, detection mechanisms relying on control
flow may generate many false alarms.

\item \textbf{Performance profiling.} Performance statistics like delay and
power consumption may increase when the inception app is running. Measuring
these metrics may provide information that aids detection. The system could
monitor parameters like CPU/GPU usage, memory access patterns, system calls,
API calls, etc., to establish a baseline, and alert if performance deviates
significantly.  However, with the unpredictable nature of
user behavior, these metrics are inherently noisy and the detection mechanism
is likely inaccurate. Moreover, the attacker's ability to adapt can introduce
further challenges, as the inception app could be carefully crafted to mimic
the performance profile of benign apps, making detection more intractable.

\item \textbf{Educating users.} Users might notice subtle anomalies but dismiss
them as bugs or glitches (see our user study in \S\ref{sec:eval}). Inception
may be detected if users are aware of this form of attacks and are alert to
minor changes in appearance or experience.  Furthermore, users can
introduce frequent customization of VR home features, e.g.,
background, to make it harder for the attacker to replicate their home
environment at a high precision.  

On the other hand, user self-detection of cyber
attacks tends to be challenging and unreliable in general. Specifically for
VR, with users accustomed to imperfect VR systems, this defense is unlikely
to be effective. Here we also note that the immersive and captivating experiences
provided by VR systems could ``overwhelm'' the users and diminish
their alertness and awareness of any subtle indicators of the 
inception attacks.  And new generations of VR systems (thorough either 
hardware or 
software advances) will only amplify the effect of overwhelmingness even
further. 
\end{packed_itemize}

\para{Defenses via Hardware.}\label{appendix:def_hw}
In addition to regular restarts, another defense technique on the hardware side is {\bf {\em regular hard resets}}.
Hard-resetting the headset wipes the device and
hence completely removes the inception app and the shell script. The adversary
would need to inject the inception app again to attack the user.  We note
that wiping the device significantly disrupts user experience, hence
this defense is quite extreme. Finally, the previously described use
of {\bf {\em a safe
bootloader and secure enclave}} is also a hardware-based defense
strategy. Although highly effective, its practical deployment is
hindered by significant cost issues, particularly for mobile devices
including VR headsets. 

\end{document}